\documentclass[aps,prd,twocolumn,groupedaddress,nofootinbib,showpacs,eqsecnum,useAms]{revtex4}

\usepackage{euscript,graphicx}
\usepackage{amsthm,dsfont,amsfonts,amsmath,amssymb}
\usepackage{euscript,color,fontenc,textcomp,relsize}
\usepackage{bm,url,float,cleveref}
\usepackage[caption=false]{subfig}

\begin{document}


\title{Gravitational waves from spinning compact binaries in hyperbolic orbits}

\author{Lorenzo De Vittori\,$^{1,}$\footnote{lorenzo@physik.uzh.ch},
 Achamveedu Gopakumar\,$^{2,}$\footnote{gopu@tifr.res.in}, 
 Anuradha Gupta\,$^2$, Philippe Jetzer\,$^1$}

\affiliation{$^1$Physik-Institut, 
Universit\"{a}t Z\"{u}rich, 8057 Z\"{u}rich, 
Switzerland\\$^2$Department of Astronomy and Astrophysics,
 Tata Institute of Fundamental Research, Mumbai 400005, India}

\begin{abstract}
  Compact binaries in hyperbolic orbits are plausible gravitational wave (GW) sources for the
upcoming and planned GW observatories.
We develop an efficient prescription to compute  post-Newtonian (PN) accurate ready-to-use 
GW polarization states for spinning compact binaries, influenced by
the dominant order spin-orbit 
interactions, in hyperbolic orbits. 
This is achieved by invoking the 
1.5PN accurate  quasi-Keplerian parameterization 
for the radial sector of the orbital dynamics.
We probe the influences of spins and 
gravitational radiation reaction on $h_{+}$ and $h_{\times}$ during the hyperbolic passage.
It turns out that both polarization states exhibit the memory effect for GWs from 
spinning compact binaries in hyperbolic orbits. 
In contrast, only cross polarization state exhibits the memory effect for GWs from
non-spinning compact binaries. Additionally,
we compute 1PN accurate amplitude corrected
GW polarization states for hyperbolic non-spinning compact binaries  
in a fully parametric manner and perform initial comparisons 
with the existing waveforms.
\end{abstract}


\maketitle

\section{Introduction}

Compact binaries in unbound orbits are 
plausible GW sources for both the ground and space based GW observatories \cite{detectors}.
These rare events are expected to occur in dense stellar environments that are present in
globular clusters and galactic nuclear clusters \cite{KGM06}.
Interestingly, 
such close encounters can, in principle, create bound binaries having very high eccentricities \cite{H72,WW79}.
For the ground based detector like the advanced LIGO \cite{H13},
the plausible  detection rates for such eccentric binaries may become 
comparable to that for  isolated compact binary coalescences, 
estimated to be between few to thousands per year \cite{OKL09}.
Very recently, it was pointed out that
electro-magnetic flares may accompany 
close encounters associated with compact binaries in hyperbolic orbits,
provided such binaries contain a neutron star \cite{DT13}.
These electro-magnetic flashes, termed as the resonant shattering flares, arise due to 
the possible crustal shattering of the neutron star during its hyperbolic passage.
The shattering develops because of the excitation of certain interface modes due to the 
extraction of orbital kinetic energy through resonant tidal coupling.
This astrophysically plausible scenario should be an interesting candidate for 
triggered GW burst searches as it involves certain electro-magnetic flares of estimated
luminosity $ \sim 10^{47} {\rm erg /s} $ \cite{DT13}.

  The investigations dealing with compact binaries in hyperbolic orbits had a chequered history
and we begin by listing papers that provided inputs required to construct the associated GW polarization
states.
The quadrupolar order gravitational radiation field associated with two non-spinning
compact objects moving in Newtonian hyperbolic orbits was 
analyzed by Turner \cite{T77}.
Its extension to the first post-Newtonian (1PN) order is available in Ref.~\cite{JS92} while invoking 
the quasi-Keplerian approach to describe 1PN accurate hyperbolic 
orbits \cite{DD85}.
Note that the 1PN accurate orbital dynamics include 
general relativity based corrections to compact binary dynamics that are accurate to  
$(\rm v/c)^2$ order beyond the Newtonian description, where $\rm v$ and $\rm c$ are the orbital 
and light speeds, respectively.
The explicit 1PN order  amplitude corrected expressions for the two GW polarization
states, $h_{+}$ and $h_{\times}$, are available in Ref.~\cite{MFV10}.
This paper employed certain generalized true anomaly parameterization, detailed in Ref.~\cite{MV08}, 
to describe 1PN accurate hyperbolic orbits. 
Additionally, there exists a number of investigations that probed various theoretical 
and observational aspects of  non-spinning compact 
binaries in hyperbolic orbits.
 This includes quadrupolar order energy and angular momentum losses
during hyperbolic encounters and its 1PN extensions \cite{WW76,BS98}.
Aspects of gravitational bremsstrahlung involving
large eccentricities and impact parameters were investigated in Ref.~\cite{KT77}.
Recently, Ref.~\cite{DJK12} obtained a
general analytic formula for the GW energy spectrum
associated with compact binaries in unbound orbits 
that generalized the parabolic limit computed in Ref.~\cite{BG10}.
The quadrupolar order GW strain amplitudes
and certain crude estimates for the expected rates of close
gravitational flybys for terrestrial GW interferometers were reported in  Ref.~\cite{C08}.
 It was argued in Ref.~\cite{RBF06} that 
GW burst signals, associated with stellar mass compact objects 
in nearly parabolic orbits around massive black hole (BH), should be
present in a LISA-type space based GW observatory data streams.
More recently, event rates for such  extreme-mass-ratio bursts and
the associated GW measurement accuracies
for the massive BH mass and spin were explored in Refs.~\cite{BG12,BG13-I, BG13-II}.

  In this paper, we obtain temporally evolving GW polarization states for spinning compact
binaries in PN accurate hyperbolic orbits. The spin effects are due to the leading order
spin-orbit interactions, as detailed in Ref.~\cite{BO75}, and  the conservative non-spinning 
orbital dynamics is 1PN accurate.
This implies that our orbital dynamics is fully 1.5PN accurate 
while considering compact binaries containing maximally spinning
BHs.
This is because 
the spin-orbit contributions to the orbital dynamics manifest at 1.5PN order  
for maximally spinning BH binaries \cite{K95}.
Additionally, we incorporate the quadrupolar order 
gravitational radiation reaction effects while computing 
$h_{+}(t)$ and $h_{\times}(t)$. The plots for 
GW polarization states having quadrupolar order amplitudes and 
PN accurate orbital evolution reveal that both polarization states
exhibit the {\it memory effect} with the inclusion of spin effects.
In contrast, only the plus polarization state
exhibits the memory effect for non-spinning compact binaries \cite{JS92,Favata09,Favata11}.
Recall that Ref.~\cite{BG86} coined 
the non-vanishing difference between the wave amplitudes  
at $t= \pm \infty$ as the {\it memory effect} while dealing with non-spinning compact binaries.
This is a {\it linear} memory effect in contrast to the non-linear memory effect present in GW polarization states
for compact binaries in bound orbits \cite{Favata11}.
The influences of orbital eccentricity, mass ratio and initial dominant spin orientation
on the observed memory effect are also probed.
We observe that the memory amplitude approaches zero as orbital eccentricity tends to unity
while time domain GW polarization states develop sharply varying features for low eccentricities.
The GW memory amplitude is larger for the cross polarization compared to the plus polarization
and weakly depends on the mass ratio. 
The amplitude of the memory effect slowly changes as we vary the 
initial orientation of the dominant spin. These changes are more visible for the plus polarization state for 
higher eccentricities.
Additionally, we  provide 1PN accurate amplitude corrected expressions for 
the two GW polarization states associated with hyperbolic spinning compact binaries in a fully parametric way.
 These expressions generalize the computations of Ref.~\cite{MFV10} that dealt with non-spinning compact
 binaries in hyperbolic orbits.
We observe that  our approach to compute  $h_{+}(t)$ and $h_{\times}(t)$
should be accurate and computationally cheaper than the one in \cite{MFV10}.
This is because of invoking Mikkola's method \cite{Mikkola}
to solve the 1PN accurate Kepler Equation for hyperbolic orbits.
 
We provide an explanation for the presence of the linear memory effect in both the polarization states 
for spinning compact binaries in hyperbolic orbits.
For this purpose, we follow the arguments that are used to explain the presence
of this effect in certain components of the far-zone metric 
associated with non-spinning compact binaries in hyperbolic orbits.
We show that these arguments ensure the presence of the linear memory effect 
in the quadrupolar order cross polarization state for non-spinning compact binaries 
in hyperbolic orbits.
This is beneficial as we can pinpoint terms that cause the effect in the case
of non-spinning compact binaries.
In contrast, we argue that  
the memory effect arises 
from the combined influences of a number of terms that are present in both polarization states associated 
with spinning compact binaries.
Invoking non-spinning compact binaries in PN accurate orbits also allow 
us to compare GW polarization states from our approach with those 
available in the literature.
Influenced by Figs.~6 to 10 in Ref.~\cite{MFV10}, we plot Newtonian, 0.5PN and 1PN  contributions 
to $h_{+}$  and $h_{\times}$  for non-spinning compact binaries in hyperbolic orbits.
A visual comparison reveals substantial differences the way Newtonian and 0.5PN contributions to 
$h_{\times}$ evolve during the hyperbolic passage in our approach and the one detailed in Ref.~\cite{MFV10}.
However, the plots in Fig.~8 of  Ref.~\cite{MFV10} for their 1PN order multipolar corrections to 
$h_{+}$  and $h_{\times}$ look qualitatively similar to our plots for the 1PN order amplitude corrections to
GW polarizations states.
 We provide a possible qualitative explanation for these differences.
Our approach indeed reproduces the temporal evolution for the real and imaginary parts of 
the time derivatives of mass and current multipole moments
and associated GW modes computed in Refs.~\cite{JS92,Favata11}.
We have invoked Fig.~8 in \cite{JS92} and Fig.~2 in \cite{Favata11} for such comparisons.
 
\begin{figure}
  \includegraphics[width=0.45\textwidth]{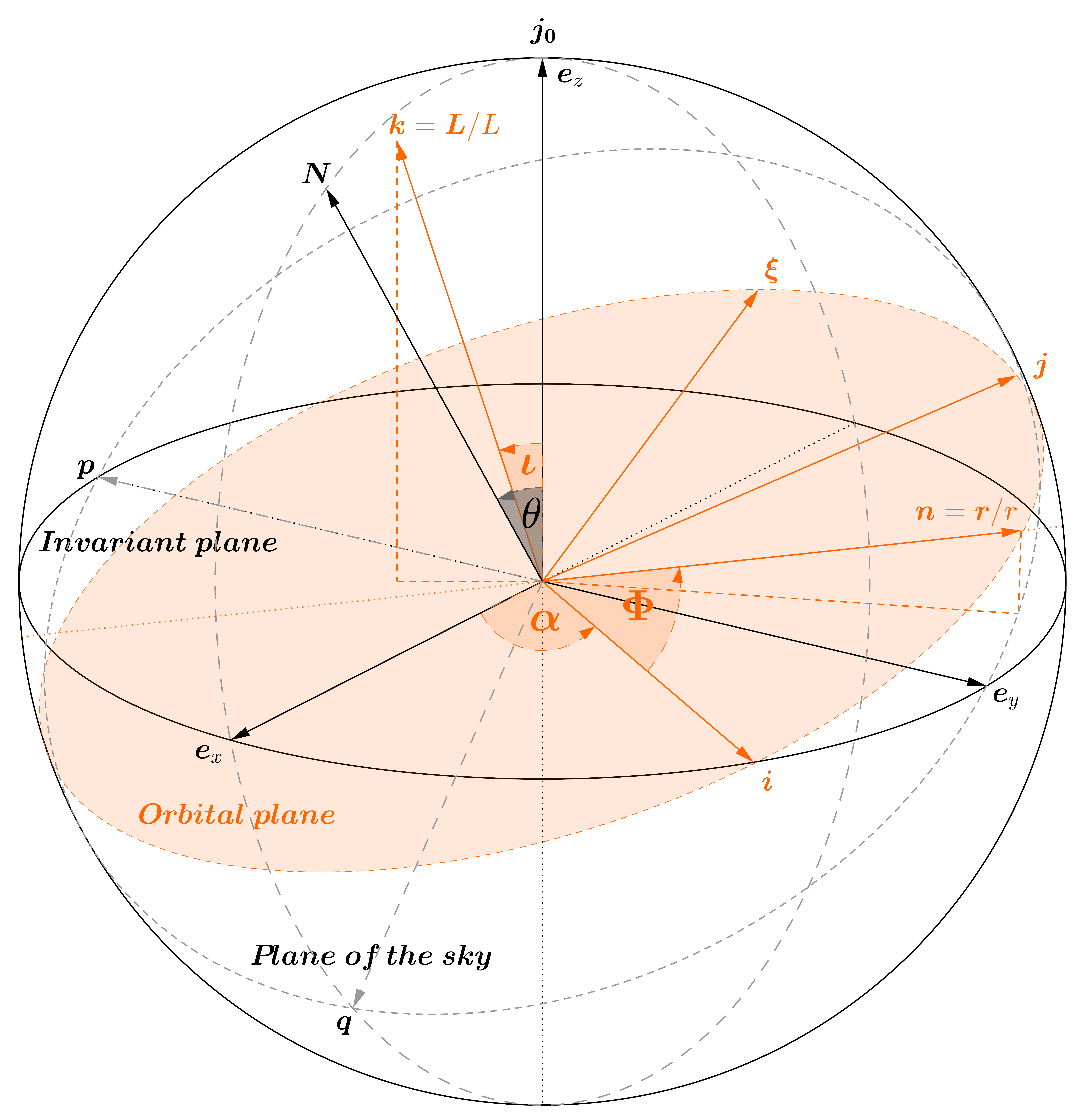}
  \caption{Various inertial and non-inertial coordinate systems that are useful to describe the dynamics of 
    spinning compact binaries and associated GWs.
    The depicted vectors $(\bm n, \bm \xi, \bm k) $ and $(\bm i, \bm j, \bm k) $  define the two non-inertial frames, namely 
    the co-moving and the orbital triads.
    The two inertial frames associated with $\bm j_0$ and $\bm N$ are also displayed, namely  
    $(\bm e_x, \bm e_y, \bm e_z) $ and $(\bm p, \bm q, \bm N) $. The orbital phase $\Phi$ of the binary 
    requires us to invoke the two vectors $\bm n$ and $\bm i$ while the orientation of 
    orbital angular momentum is specified by the two angles $\iota$ and $\alpha$ present in the $\bm j_0$ based 
    inertial frame. 
    The $(\bm p, \bm q, \bm N) $ frame is essentialy specified by the angle $\theta$ between $\bm N$ and $\bm j_0$.
  It should be noted that the orbital plane $(\bm i, \bm j, \bm k) $ precesses around $\bm j_0$ due to spin-orbit coupling.}
  \label{fig:frames}
\end{figure}

The paper is organized in the following way. In Sec.~\ref{sec:spinning} we 
present our approach to obtain temporally evolving GW polarization states 
for spinning compact binaries in hyperbolic orbits during their close encounters.
We focus on non-spinning compact binaries in Sec.~\ref{sec:nonspinning} influenced by  Ref.~\cite{MFV10}
and visually compare the evolution of $h_{\times}$ and $h_{+}$ in these two approaches.
A brief summary, possible consequences and extensions are listed in Sec.~\ref{sec:conclusions}.
\newpage

\section{ Waveforms for spinning compact binaries in hyperbolic orbits}\label{sec:spinning}

  We begin by listing the explicit expressions for the quadrupolar order GW polarization states for spinning 
compact binaries  moving in non-circular 
orbits.
\onecolumngrid
\begin{align}
  h_+|_Q(t) =\;&\frac{2\,G\,m\, \eta }{c^4\,R}\; \bigg\{\left(\dot{r}^2 - z\right)\bigg[(\sin{\alpha}\cos{\Phi}+\cos{\iota}\cos{\alpha}\sin{\Phi})^2-(C_{\theta}(\cos{\Phi}\cos{\alpha}-\sin{\alpha}\cos{\iota}\sin{\Phi})-S_{\theta}\sin{\iota}\sin{\Phi})^2\bigg] \nonumber\\
  &+\;r^2 \,\dot{\Phi}^2 \,\bigg[(\cos{\alpha}\cos{\iota}\cos{\Phi}-\sin{\alpha}\sin{\Phi})^2-(C_{\theta}\sin{\alpha}\cos{\iota}\cos{\Phi}+C_{\theta}\cos{\alpha}\sin{\Phi}-S_{\theta}\sin{\iota}\cos{\Phi})^2\bigg] \nonumber\\
  &-\;r \,\dot{r} \,\dot{\Phi} \,\bigg[ \cos^2\alpha \cos\Phi \sin\Phi (\cos^2\iota + C_{\theta}^2) +\cos\alpha (\cos^2\Phi - \sin^2\Phi) \left(\cos\iota \sin\alpha (1 + C_{\theta}^2) + C_{\theta} S_{\theta} \sin\iota \right) \nonumber\\
  &-\cos\Phi \sin\Phi \left((1 + \cos^2\iota \,C_{\theta}^2) \sin^2\alpha + 2 C_{\theta} S_{\theta} \cos\iota \sin\alpha \sin\iota + \sin^2\iota S_{\theta}^2 \right) \bigg]\bigg\}~,\label{eq:polstateP}\\
  h_{\times}|_Q(t) =\;&\frac{4\,G\,m\,\eta}{c^4\,R}\; \bigg\{\left(\dot{r}^2  - z\right)\bigg[(\sin{\alpha}\cos{\Phi}+\cos{\iota}\cos{\alpha}\sin{\Phi})(C_{\theta}(\cos{\Phi}\cos{\alpha}-\sin{\alpha}\cos{\iota}\sin{\Phi})-S_{\theta}\sin{\iota}\sin{\Phi})\bigg] \nonumber\\
  &-\;r^2 \,\dot{\Phi}^2 \,\bigg[(\cos{\iota}\cos{\Phi}\cos{\alpha}-\sin{\alpha}\sin{\Phi})(C_{\theta}(\cos{\Phi}\sin{\alpha}\cos{\iota}+\cos{\alpha}\sin{\Phi})+S_{\theta}\sin{\Phi}\sin{\iota})\bigg] \nonumber\\
  &+\;r \,\dot{r} \,\dot{\Phi} \,\bigg[ \cos^2\alpha \cos\iota C_{\theta} (\sin^2\Phi-\cos^2\Phi) + \sin\alpha (\cos^2\Phi - \sin^2\Phi) (\cos\iota C_{\theta} \sin\alpha + \sin\iota S_{\theta}) \nonumber\\
  &+\; 2 \cos\alpha \cos\Phi \sin\Phi \,\left((1 + \cos^2\iota) C_{\theta} \sin\alpha + \cos\iota \sin\iota S_{\theta}\right) \bigg]\bigg\}~,\label{eq:polstateC}
\end{align}
\twocolumngrid
\noindent where $R, S_{\theta}$ and $ C_{\theta}$ stand 
for the radial distance to the binary, $\sin \theta$ and $\cos \theta$, respectively, and $z = Gm/r$,
where $G$ denotes the gravitational constant.
We would like to warn the reader that at few places the character
$z$ is also associated with the unit vector $\bm{z}$, the $z$-axis
of our Cartesian coordinate system and the $z$-component of unit
vectors like $k_z$, as commonly used in the literature.
The angle $\theta$ provides the angle between the line of sight vector $\bm{N}$ 
and $\bm{j}_0$, the unit vector along the direction of the total angular momentum at the initial
epoch (see Fig. \ref{fig:frames}).
The above expressions are provided in 
an inertial frame where $\bm{j}_0$ points along the  $z$-axis 
and where the dynamical angular variable $\Phi $
measures the orbital phase from the line of nodes
that coincides with the unit vector $\bm i$
 in a plane perpendicular to $\bm L$.
Additionally, $r$ and $ \dot r $ denote the radial orbital separation and its time derivative, respectively,
while $\dot \Phi = \text{d} \Phi/\text{d}t$.
The angles $\iota$ and $\alpha$ specify the orientation of the orbital
angular momentum $\bm{L}$ in the $\bm{j}_0$ based inertial frame.
In particular, $\iota$ specifies the angle between the orbital
angular momentum $\bm{L}$ and the $z$-axis of the inertial frame 
while 
$\alpha$  denotes the angle between the $y$ axis and the projection of 
$\bm{L}$ onto the $x-y$ plane of the $\bm{j}_0$ based inertial frame.
The notations $m$ and $\eta$ stand for $m=m_1+m_2$ and
$\eta= m_1m_2/m^2$. In what follows, we sketch briefly how we 
obtained the above expressions for $h_+|_Q(t)$ and $h_{\times}|_Q(t)$.\\

It is customary to compute PN accurate expressions for GW polarization states from the following formulae 
that contain $h_{ij}^\text{TT}$, the transverse--traceless part of the radiation field:
\begin{subequations}
\label{eq:definition_hp_hx}
\begin{align}
h_{+} &= \frac{1}{2} \left(p_i p_j - q_i q_j \right) h_{ij}^\text{TT}
\,, \\
h_{\times} &= \frac{1}{2} \left(p_i q_j + p_j q_i \right) h_{ij}^\text{TT}
\,,
\end{align}
\end{subequations}
where the vectors $\bm{p} $ and $\bm{q} $ form an orthonormal triad with the 
line--of--sight unit vector
$\bm{N}$  such that 
$ \bm{p} = \bm{N} \times \bm{j}_0 $ and 
$\bm{q} = \bm{N} \times \bm{p}$ \cite{BIWW96}.
To compute the quadrupolar order GW polarization states, we require 
the expression for $h_{ij}^\text{TT}$ that arises from the time varying 
Newtonian order quadrupole moment of the binary.
The quadrupolar order contribution to $h_{ij}^\text{TT}$ reads
\begin{equation}\label{eq:definition_h_newton}
  h_{ij}^\text{TT} \big|_{\text Q} = \frac{4 G \mu }{ c^4 R} {\cal P}_{kmij}(\bm{N})\left( \text{v}_{km} - \frac{Gm}{r} n_{km} \right)~,
\end{equation}
where ${\cal P}_{kmij}(\bm{N})$ is the transverse traceless
projection operator projecting vectors onto the plane orthogonal to $\bm{N}$ and
$\mu $ being the reduced mass ($\mu=m_1\,m_2/m$).
Additionally, $\text{v}_{ij} $ and $n_{ij}$ stand for $\text{v}_i\text{v}_j$ and $ n_i n_j $,
where $n_i$ and $\text{v}_i$ denote the components of $\bm{n} = \bm{r}/r$ and 
the velocity vector $\textbf{v} = \text{d} \bm{r} / \text{d}t$.
It should be noted that the dynamical variables appearing in eqs.~(2.1) and (2.2)
follow 1.5PN-accurate orbital evolution though we use Newtonian order expression for $h_{ij}^{TT}$.
This is influenced by the restricted PN waveforms for quasi-circular inspiral
where the orbital frequency and phase follow PN-accurate evolution
though the expressions for $h_+$ and $h_{\times}$ arise from the quadrupolar
order  $h_{ij}^{TT}$.
This leads to the following symbolic expressions for 
$h_+|_Q(t) $ and $h_{\times}|_Q(t) $
\begin{subequations}
  \label{eq:h_plus_and_h_cross_in_n_and_v}
  \begin{align}
    \label{eq_h+}
    h_{+} &= \frac{2 G \mu}{c^4 R} \bigg\{ \!(\bm{p} . \textbf{v} )^2 \!- (\bm{q} . \textbf{v} )^2 - z\! \left[ ( \bm{p} . \bm{n} )^2 \!- ( \bm{q} . \bm{n} )^2 \right] \!\bigg\}, \\
    \label{eq_hx}
    h_{\times} &= \frac{4 G \mu}{c^4 R}\bigg\{(\bm{p} . \textbf{v}) (\bm{q} . \textbf{v}) - z\,(\bm{p} . \bm{n}) (\bm{q} . \bm{n}) \bigg\}~.
  \end{align}
\end{subequations}
It is convenient to evaluate the above dot products by expressing the vectors 
$\bm{n}, \textbf{v}, \bm p, \bm q$ and $\bm N$ in a co-moving triad
$( {\bm{n}},{\bm{\xi}} = \bm{k} \times \bm{n},{\bm{k}} )$, where 
$\bm{k}$ is the unit vector along $\bm{L}$.
It is easy to deduce that the components of these three unit vectors 
in the $\bm{j}_0$ based inertial frame are specified by the usual
three Eulerian angles $\Phi, \alpha$ and  $\iota$ \cite{DS88}. In our convention,
the inertial frame components of ${\bm{n}},{\bm{\xi}} $ and $\bm{k}$ are 
given by
\begin{subequations}
  \begin{align}
  \bm{n} =\;  &(\cos{\alpha} \cos{\Phi} - \cos{\iota} \sin{\alpha} \sin{\Phi})\;\bm{x}\,+\\
	      &(\sin{\alpha} \cos{\Phi} + \cos{\iota} \cos{\alpha} \sin{\Phi})\;\bm{y}+(\sin{\iota} \sin{\Phi})\;\bm{z}~,\nonumber\\
  \bm{\xi} =\;&(-\cos{\alpha} \sin{\Phi}- \sin{\alpha} \cos{\iota} \cos{\Phi})\;\bm{x}\;+\\
	      &(\cos{\iota} \cos{\Phi} \cos{\alpha} - \sin{\alpha} \sin{\Phi})\;\bm{y}+(\sin{\iota} \cos{\Phi})\;\bm{z}~,\nonumber\\
  \bm{k} =\;  &\sin{\alpha} \sin{\iota}\;\bm{x}-\cos{\alpha} \sin{\iota}\;\bm{y}+\cos{\iota}\;\bm{z}~.\label{eq:nxk}
\end{align}
\end{subequations}
Invoking three rotations that involve the above three Eulerian angles, 
it is straightforward to express the vectors that appear in Eqs.~(\ref{eq:h_plus_and_h_cross_in_n_and_v})
in the $( \bm n, \bm \xi, \bm k)$ co-moving triad.
The resulting expressions read
\begin{subequations}
\label{eq:p_q_r_v}
\begin{align}
\label{eq:r_comov}
\bm{r} = &r\bm{n}~,  \\  \label{eq:v_comonv}
\textbf{v} = &\dot r\, \bm n +r\, \biggl (  \frac{ d \Phi}{dt} +   \frac{ d \alpha}{dt}\,\cos \iota \biggr ) \, \bm \xi\;+\\
        &r\, \biggl (  \frac{ d \iota }{dt}\, \sin  \Phi - \sin \iota \, \cos \Phi\,  \frac{ d \alpha}{dt} \biggr )\, \bm k~,\nonumber\\
\bm p =\; &(-\sin\alpha\, \cos\Phi - \cos\iota\,\cos\alpha\, \sin\Phi)\, \bm n\\
	  &+(\sin\alpha\, \sin\Phi - \cos\iota\, \cos\alpha\, \cos\Phi)\, \bm \xi\nonumber\\
	  &+\cos\alpha\, \sin\iota\, \bm k~,\nonumber\\
\bm q =\; &(\cos\alpha\, \cos\Phi\, \cos\theta - \cos\iota\, \sin\alpha\, \sin\Phi\, \cos\theta\\
	  &- \sin\iota\, \sin\Phi\, \sin\theta)\, \bm n\, -(\cos\alpha\, \sin\Phi\, \cos\theta \nonumber\\
	  &+ \sin\alpha\, \cos\iota\, \cos\Phi\, \cos\theta + \sin\iota\, \cos\Phi\, \sin\theta)\, \bm \xi\nonumber\\
	  &+(\sin\alpha\, \sin\iota\, \cos\theta - \cos\iota\, \sin\theta)\, \bm k~,\nonumber\\
\bm N =\; &(\cos\alpha\, \cos\Phi\, \sin\theta - \cos\iota\, \sin\alpha\, \sin\Phi\, \sin\theta\\
	  &+\sin\iota\, \sin\Phi\, \cos\theta)\, \bm n\, -(\cos\alpha\, \sin\Phi\, \sin\theta \nonumber\\
	  &+\sin\alpha\, \cos\iota\, \cos\Phi\, \sin\theta - \sin\iota\, \cos\Phi\, \cos\theta )\, \bm \xi\nonumber\\
	  &+(\sin\alpha\, \sin\iota\, \sin\theta + \cos\iota\, \cos\theta)\, \bm k~.\nonumber
\end{align}
\end{subequations}
To obtain the above expressions, we invoked the definitions for $\bm p$, $\bm q$
and let $\bm j_0$, $\bm N$ have the following components in the inertial frame:
$\bm j_0= (0,0,1)$ and $\bm N = (\sin \theta, 0, \cos \theta)$.
It is not difficult to verify that an explicit evaluation of
Eqs.~(\ref{eq:h_plus_and_h_cross_in_n_and_v}) for $h_+|_Q(t) $
and $h_{\times}|_Q(t) $ while employing the above expressions for 
$\bm n, \textbf{v}, \bm p, \bm q$ and $\bm N$ results in Eqs.~(2.1) and (2.2).\\

  We obtain temporally evolving $h_+|_Q(t) $ and $h_{\times}|_Q(t) $ for spinning compact binaries
in hyperbolic orbits by specifying  how 
$r, \dot r, \iota, \alpha, \Phi$ and $\dot \Phi$ evolve in time along PN accurate hyperbolic orbits.
The radial part of the dynamics is tackled in a parametric manner invoking $v$, a real variable along the orbit,
the time eccentricity $e_{\rm t}$ and the mean motion $\bar n$ associated with PN accurate hyperbolic 
orbits of Ref.~\cite{DD85}. 
The 1.5PN accurate parametric expressions for $r $ and $\dot r$, adapted from Ref.~\cite{GS11}, read 
\begin{subequations}
\label{eq:r}
\begin{align}
  \label{eq:r-a}
  r =\; &\frac{Gm}{c^2}\frac{1}{\bar \xi^{2/3}}\,\bigg\{\;e_{\rm t} \cosh{v} -1\; - \\
	   &\bar \xi^{2/3}\;\frac{e_{\rm t}\cosh{v}\,(6-7\eta)+18-2\eta}{6}\;+\bar \xi\;\frac{\Sigma}{\sqrt{e_{\rm t}^2-1}}\;\bigg\}\nonumber~,\\
\label{eq:rdot}	   
  \dot{r} = \;&\bar \xi^{1/3}\frac{c\,e_{\rm t}\,\sinh{v}}{e_{\rm t}\cosh{v}-1}\,\bigg\{1-\bar \xi^{2/3}\,\frac{6-7\eta}{6}\bigg\}~,
  \end{align}
\end{subequations}
where $\bar \xi = Gm\bar n/c^3$ and $\Sigma$ terms are due to the spin-orbit interactions.
The expression for $\Sigma$ is defined as 
\begin{equation}
  \Sigma=\delta_1\chi_1q\,(\bm{k}\cdot\bm{s}_1)+\frac{\delta_2\chi_2}{q}\,(\bm{k}\cdot\bm{s}_2)~,
\end{equation}
where $\delta_1 = \eta/2 + 3/4 (1-\sqrt{1-4\eta})$, and $\delta_2 = \eta/2 + 3/4(1+\sqrt{1-4\eta})$ while 
$q= m_1/m_2$ with $m_1 \geq m_2$.
The dot products define the misalignments between $\bm L$ and the two spins $\bm S_1 $ and $\bm S_2$ 
while $\chi_{1,2}$ stand for the two Kerr parameters such that $\bm S_1 = G\, m_1^2 \, \chi_1\,\bm s_1/c $ and 
$\bm S_2 = G\, m_2^2 \,\chi_2\, \bm s_2/c $.
The temporal evolution for $r $ and $\dot r$ are obtained by solving the following hyperbolic version of 
the classical Kepler Equation 
\begin{align}\label{eq:bar_n}
 \bar n\;(t-t_0) &= e_{\rm t} \sinh v - v\,,
\end{align}
where $t_0$ is certain initial epoch.
In the present investigation, we invoke an accurate and efficient numerical procedure, namely Mikkola's 
approach \cite{Mikkola}, to obtain $v(t)$ from the above transcendental equation.

Let us note that the above three equations, namely Eqs.~(2.8a), (2.8b) and (2.10),
are adapted from Ref.~\cite{GS11} that obtained Keplerian type parametric
solution to the radial sector  of spinning compact binary dynamics  in eccentric 
orbits.  We begin by listing relevant equations that describe the above
mentioned 1.5PN accurate Keplerian type parametric solution: 
\begin{subequations}
\begin{align}
  r &= a_{\rm r} \;(1-e_{\rm r} \cos u)~,\\
  l &= n\;(t-t_0) = u - e_{\rm t} \sin u~, 
  \end{align}
\end{subequations}
where $u$ and $l$ stand for the eccentric and mean anomalies. 
In what follows, we employ $l$ to explore various aspects of our time domain 
GW polarization states as $l$ essentially represents the scaled coordinate time.
The orbital parameters $a_{\rm r}$ and $n$  are the PN 
extensions of the semi-major axis and the mean motion associated with the Keplerian parametric solution to the 
Newtonian orbital dynamics. Additionally, the radial part of the 
PN accurate Keplerian type parameterization involves two eccentricities, namely
$e_{\rm r}$ and $e_{\rm t}$ \cite{DD85}.
These orbital elements are explicit functions of the reduced orbital energy $E$, 
reduced angular momentum $L$, the symmetric mass ratio $\eta$, the two Kerr parameters
and the spin-orbit misalignments.
To 1.5PN order, these parameters are given by 
\begin{subequations}\label{eq:ellparam}
\begin{align}
  a_{\rm r} =\, &\frac{Gm}{-2E}\bigg\{1-\frac{2E}{4c^2}(\eta-7)-\frac{2E}{c^3}\frac{Gm\,\Sigma}{L}\bigg\}~,\\
  e_{\rm t}^2 =\, &1 + \frac{2EL^2}{G^2m^2}-\frac{2E}{c^2}\,\bigg(\;2\eta-2\;+\\
		  &\frac{2EL^2}{G^2m^2}\frac{7\eta-17}{4}\bigg)+\frac{4E}{c^3}\frac{Gm\,\Sigma}{L}~,\nonumber\\
  e_{\rm r}^2 =\, &1 + \frac{2EL^2}{G^2m^2}-\frac{2E}{c^2}\,\bigg(\,6-\eta\;+\\
		&\frac{2EL^2}{G^2m^2}\frac{15-5\eta}{4}\bigg)+\frac{8E}{c^3}\frac{Gm\,\Sigma}{L}\left(1+\frac{EL^2}{G^2m^2}\right)~,\nonumber\\
  \xi = &\left(\frac{-2E}{c^2}\right)^{3/2}\left\{1-\frac{2E}{c^2}(\eta-15)\frac18\right\}~\,,
\end{align}
\end{subequations}
where $\xi$ stands for $Gmn/c^3$.
The structure of the two eccentricities indicate that it should be possible to express $e_{\rm r}$ in terms of 
$e_{\rm t}$ as a PN series employing $\xi$ and the spin parameters (this holds true for $a_{\rm r}$).
The resulting expressions for $e_{\rm r}$ and $a_{\rm r}$ read
\begin{subequations}\label{eq:relellparam}
\begin{align}
  e_{\rm r} &= e_{\rm t} \left\{1+\xi^{2/3}\,\frac{8-3\eta}{2}-\xi\,\frac{\Sigma}{\sqrt{1-e_{\rm t}^2}}\right\}~,\\
  a_{\rm r} &= \frac{Gm}{c^2}\frac{1}{\xi^{2/3}} \left\{1-\xi^{2/3}\frac{9-\eta}{3}+\xi\frac{\Sigma}{\sqrt{1-e_{\rm t}^2}}\,\right\}~.
\end{align}
\end{subequations}
The 1.5PN accurate expressions for $r$ and $ \dot r$ in terms of $\bar n$, $e_{\rm t}$ and $v$ are 
obtained with the help of the following steps. 
First, we obtain explicit 1.5PN accurate expression for $r= a_{\rm r} ( 1- e_{\rm r}\, \cos u)$ in terms of 
$n$, $e_{\rm t}$ and $u$ with the help of Eqs.~(\ref{eq:relellparam}).
This leads to 
\begin{align}
  r  =\; &\frac{Gm}{c^2}\frac{1}{\xi^{2/3}}\bigg\{1- e_{\rm t}\cos u - \frac{\xi^{2/3}}{6}\bigg(\,18-2\eta\\
	 &+(6-7\eta)\,e_{\rm t}\cos u\bigg)+\xi\,\frac{\Sigma}{\sqrt{1-e_{\rm t}^2}}\,\bigg\}~.\nonumber
  \end{align}
To obtain its hyperbolic counterpart,  
we let $u = \imath\, v$ and $n = -\imath\, \bar n$, where $ \imath = \sqrt{-1}$,
by invoking the arguments in Ref.~\cite{DD85}. This analytic continuation in $E$
from $E <0$ to $E > 0$ essentially works as all orbital parameters that are analytic near $E =0$.
We employ similar arguments to obtain $\dot r (v,\bar n, e_{\rm t}) $, given by Eq.~(2.8b), 
from its eccentric version computed in terms of $u, n$ and $ e_{\rm t}$.

  With the help of the above arguments and Ref.~\cite{GS11},
we extract the following 1.5PN accurate $\dot \Phi$ expression for spinning compact binaries in hyperbolic 
orbits 
\begin{align}\label{eq:dot_phi}
  \dot \Phi =&\, \frac{\bar n\,\sqrt{e_{\rm t}^2-1}}{(e_{\rm t}\cosh v -1)^2}\bigg\{1+\bar \xi^{2/3}\left(\frac{4-\eta}{e_{\rm t}\cosh v -1}+\frac{\eta-1}{e_{\rm t}^2-1}\right)\nonumber\\
	 &-\bar \xi \,\frac{\Sigma}{\sqrt{e_{\rm t}^2-1}} \left(
 \frac{1}{e_{\rm t}\cosh v -1}+\frac{1}{e_{\rm t}^2-1}\right)\!\bigg\}-\cos\iota\;\dot{\alpha}\,,
\end{align}
where the differential equation for $\alpha$ arises from the precessional equation for $\bm{k}$.
The above differential equation for $\Phi$ is also adapted from its eccentric counterpart, given by Eq.~(B2) in Ref.~\cite{GS11}.
To derive Eq.~(B2) in \cite{GS11}, one starts from 
the expression for $\textbf{v}$ in the co-moving triad
$(\bm{n}, \bm \xi, \bm{k})$ as given by Eq.~(2.7b).
This leads to the following 1.5PN accurate expression for $\textbf{v} \cdot \textbf{v}$, namely
\begin{equation}
  \text{v}^2 = \dot r ^2 + r^2(\dot{\Phi}^2+2\dot{\Phi}\dot{\alpha}\cos\iota)~,
\end{equation}
where we have neglected the $\mathcal{O}(1/c^6)$ order $\dot{\alpha}^2$, $\dot{\iota}^2$ and $\dot{\alpha}\dot{\iota}$ terms.
The next step is based on the fact that the square of the orbital velocity, extractable from the
fully 1.5PN order Hamiltonian or orbital energy, does not contain any spin dependent terms.
This is attributable to the employed gauge and the spin supplementrary condition in Ref.~\cite{GS11}.
This is why one obtains Newtonian order relation, namely  $\text{v}^2 = 2E+2Gm/r$,
while computing  $\text{v}^2$ from a Hamiltonian that only contains Newtonian and 1.5PN order 
spin-orbit contributions. This statement  may be verified by inspecting the 
 Eqs.~(1), (5), (10) and (39) of \cite{GS11}.
Therefore, 
the  $\text{v}^2$ expression that arise from a fully 1.5PN accurate orbital energy for spinning 
binaries in general orbits will not explicitly contain any spin-orbit contributions
similar to the parametric equation for $\dot r$, given by Eq.~(2.8b).
The PN-accurate expression for $\text{v}^2$ 
is given by
\begin{align}
  \text{v}^2 =&\;2E+2\frac{Gm}{r} + \frac{1}{c^2} \, \biggl \{ \left(9\,\eta-3 \right) {E}^{2}\,+\\
	      &\frac{Gm}{r} (14\eta -12)E+ {\frac {{G}^{2}{m}^{2}}{{r}^{2}}}(5\eta-10)+{\frac {G m \eta}{{r}^{3}}}L^2 \biggr \}~,\nonumber
\end{align}
The expression for $\dot \Phi^2 $ follows by equating 
the above two expressions for $\text{v}^2$.
This leads to PN accurate expression for $\dot \Phi $, as given by Eq.~(B2) in \cite{GS11}, where
the spin-orbit contributions arise from the PN accurate expression for $r$.
We obtain our Eq.~(2.15) for $\dot \Phi$ 
with the help of earlier mentioned analytic continuation after expressing 
$(-2E)$ and $L$ in terms of $n, e_t$ and $u$.
For easy visualization, it is convenient to characterize our hyperbolic binaries in terms of an impact parameter $b$
such that $b\, \text{v}_{\infty} = | \bm r \times \textbf{ v} |$ when $ |\bm r | \rightarrow \infty$ and where 
$\text{v}_{\infty}$ stands for the relative velocity at infinity. 
We characterize our hyperbolic binaries using the following 
1PN accurate expression for $b$ in terms of $\bar \xi$ and $e_t$ 
\begin{equation}
  b = \frac{Gm}{c^2}\frac{\sqrt{e_{\rm t}^2-1}}{\bar \xi^{2/3}}\left\{1-\bar \xi^{2/3}\left(\frac{\eta-1}{e_{\rm t}^2-1}+\frac{7\eta-6}{6}\right)\right\}~.\\
\end{equation}

 The temporal evolution for $\alpha$ and $\iota$, as expected,
requires us to solve the precessional equation for $\bm{k}$ in the 
$\bm{j_0}$ based inertial frame and this is clearly due to the
Eq.~(2.6c) for $\bm{k}$. In practice, we numerically solve
coupled precessional equations for $\bm{s_1}, \bm{s_2}$ and $\bm{k}$ 
as the differential equation for $\bm{k}$ arises from the relation 
$\dot{\bm{k}} = -( S_1\, \dot{\bm{s}_1} + S_2\, \dot{\bm{s}_2})/L $.
This equation, as expected, arises from the conservation of the total angular momentum
and the magnitude of the orbital angular momentum during the precessional timescale.
The relevant equations that incorporate the dominant order spin-orbit coupling for binaries 
in hyperbolic orbits are given by
\begin{subequations}\label{eq:prec}
  \begin{align}
    \dot{\bm{s_1}}  &= \frac{c^3}{Gm}\frac{\bar \xi^{5/3}\sqrt{e_{\rm t}^2-1}}{(e_{\rm t}\cosh v-1)^3}\;\delta_1\;\bm{k}\times\bm{s}_1~,\\
    \dot{\bm{s_2}}  &= \frac{c^3}{Gm}\frac{\bar \xi^{5/3}\sqrt{e_{\rm t}^2-1}}{(e_{\rm t}\cosh v-1)^3}\;\delta_2\;\bm{k}\times\bm{s}_2~,\\
    \dot{\bm{k}}    &= \frac{c^3}{Gm}\frac{\bar \xi^2\;(\delta_1\chi_1q\;\bm{s}_1+\delta_2\chi_2/q\;\bm{s}_2)\times\bm{k}}{(e_{\rm t}\cosh v-1)^3}~.\label{eq:prec_k}
  \end{align}
\end{subequations}
The expressions for $\dot{\bm{s_1}}$ and $\dot{\bm{s_2}}$ are adapted from 
Ref.~\cite{BO75}
while invoking the Newtonian accurate $L$ for hyperbolic orbits and 
the  Newtonian version of our Eqs.~(\ref{eq:r}) for $r$.
Additionally, we employ the following expression for $\dot \alpha$
that appear in the differential equation for $\Phi$.
This equation arises from
 the Cartesian components of $\bm{k}$ in the $\bm{j_0}$ based inertial frame
and Eq.~(2.19c) for $\dot{\bm{k}}$ :
\begin{align}
  \dot \alpha &= \frac{k_x \dot{k}_y - k_y \dot{k}_x}{k_x^2 + k_y^2}\,.
  \end{align}
It is possible to incorporate numerically the effects of GW emission during the hyperbolic encounter.
This is achieved by solving the following 2.5PN order coupled differential equations for $\bar n$ and $e_{\rm t}$:
\begin{subequations}\label{eq:radiation_reaction}
  \begin{align}
    \frac{d\bar n}{dt}  = &-\frac{c^6}{G^2m^2}\frac{\bar \xi^{11/3}\;8\;\eta}{5\;\beta^7}\;\times\\
			  &\left[-49\beta^2-32\beta^3+35(e_{\rm t}^2-1)\beta-6\beta^4+9e_{\rm t}^2\beta^2\right]~,\nonumber\\
    \frac{de_{\rm t}}{dt}  = &-\frac{c^3}{Gm}\frac{\bar \xi^{8/3}\;8\;\eta\;(e_{\rm t}^2-1)}{15\;\beta^7\;e_{\rm t}}\;\times\\
			     &\left[-49\beta^2-17\beta^3+35(e_{\rm t}^2-1)\beta-3\beta^4+9e_{\rm t}^2\beta^2\right]~,\nonumber
  \end{align}
\end{subequations}
where, for simplicity, we write $\beta=e_{\rm t}\cosh v -1$.
The derivation of the above two differential equations is adapted from Eqs.~(63) in Ref.~\cite{DGI04}
and requires 2.5PN contributions to the relative acceleration.

 We are now in a position to obtain $h_+|_Q(t) $ and $h_{\times}|_Q(t) $ for spinning compact binaries
moving in hyperbolic orbits influenced by GW emission.
The idea is to numerically obtain  the temporal evolution for $r, \dot r, \Phi, \dot \Phi, \iota, \alpha, \bar n$ and 
$e_{\rm t}$ and impose these variations in the expression for $h_+|_Q(t) $ and $h_{\times}|_Q(t) $, given by Eqs.~(2.1) and (2.2).
We begin by specifying the initial binary configuration 
in terms of $m_1, m_2, \chi_1, \chi_2, \bar n$ and $e_{\rm t}$.
It is possible to specify the initial 
 orientations of $\bm{s_1}, \bm{s_2}$ and $\bm{k}$ in the 
$\bm{j_0}$ based inertial frame 
by freely  choosing four angles $(\theta_1^i,\theta_2^i)$ and $(\phi_1^i,\phi_2^i)$.
These four angles provide the six Cartesian components of $\bm{s_1}$ and $\bm{s_2}$ 
at the initial epoch and, in general, these components are 
\begin{subequations}
  \begin{align}
    \bm{s_1}  &= (\sin\theta_1\cos\phi_1,\sin\theta_1\sin\phi_1,\cos\theta_1)~,\\
    \bm{s_2}  &= (\sin\theta_2\cos\phi_2,\sin\theta_2\sin\phi_2,\cos\theta_2)~.
  \end{align}
\end{subequations}
The values of $\iota$ and $\alpha$ that specify the initial orientation of $\bm{k}$ are not freely 
chosen. These initial estimates are obtained by noting that the initial $x$ and $y$ components of $\bm j$ 
should be zero as we let $\bm j$ to point along the $z$-axis at the initial epoch.
This leads to the following expressions for $k_x$ and $k_y$ at $t=0$
\begin{subequations}
  \begin{align}
    k_x|_{t=0}  =\; &\bar \xi_0^{1/3} \frac{X_1^2\chi_1\sin\theta_1^i\cos\phi_1^i+X_2^2\chi_2\sin\theta_2^i\cos\phi_2^i}{\eta\;\sqrt{e_{\rm t}^2-1}}~,\\
    k_y|_{t=0}  =\; &\bar \xi_0^{1/3} \frac{X_1^2\chi_1\sin\theta_1^i\sin\phi_1^i+X_2^2\chi_2\sin\theta_2^i\sin\phi_2^i}{\eta\;\sqrt{e_{\rm t}^2-1}}~,
  \end{align}
\end{subequations}
where $X_1= m_1/m$ and  $X_2= m_2/m$ while $\bar \xi_0$ denotes the initial value for $\bar \xi$.
The initial estimates for $\iota$ and $\alpha$ is obtained by equating the above expressions to 
$\sin \alpha\, \sin \iota$ and $-\cos \alpha\, \sin \iota$.
However, we usually extract values of $\alpha$ and $\iota$ during the numerical interaction 
with the help of the three Cartesian components of $\bm k$, namely 
$\alpha\!=\!-\tan^{-1}(k_x/k_y)$ and $\iota=\cos^{-1}(k_z)$.
We impose the phasing angle $\Phi$ to vanish at periastron time, i.e. $\Phi(0)=0$.
 
We begin the numerical implementation of $h_+|_Q(t) $ and $h_{\times}|_Q(t) $ by obtaining  $v(t)$ 
and this involves solving Eq.~(2.10) via the Mikkola's method.
The resulting $v(t)$ is imposed on Eqs.~(\ref{eq:r}) for $r(v)$ and $\dot r(v)$ to obtain 
1.5PN accurate $r(t)$ and $\dot r(t)$ for our hyperbolic binary configuration.
The next step involves numerically integrating simultaneously the above listed differential equations for 
$\bm{s_1}, \bm{s_2}, \bm{k}, \Phi, \bar n$ and 
$e_{\rm t} $. This is achieved by invoking  
twelve differential equations that include differential equations 
for the nine Cartesian components of $\bm{s_1}, \bm{s_2} $ and $ \bm{k}$ in the $\bm{j_0}$ based inertial frame.
In practice, we use the mean anomaly $l$ rather than the coordinate time $t$ while numerically 
tackling these differential equations and 
the transcendental equation (2.10). The change of variable is performed by noting that $\text{d}l = \bar n \, \text{d}t$.
In what follows, we display $h_+|_Q(l) $ and $h_{\times}|_Q(l) $ resulting from such an implementation and 
explore various features.

\onecolumngrid

  \begin{figure}[h]
    \includegraphics[width=.77\textwidth]{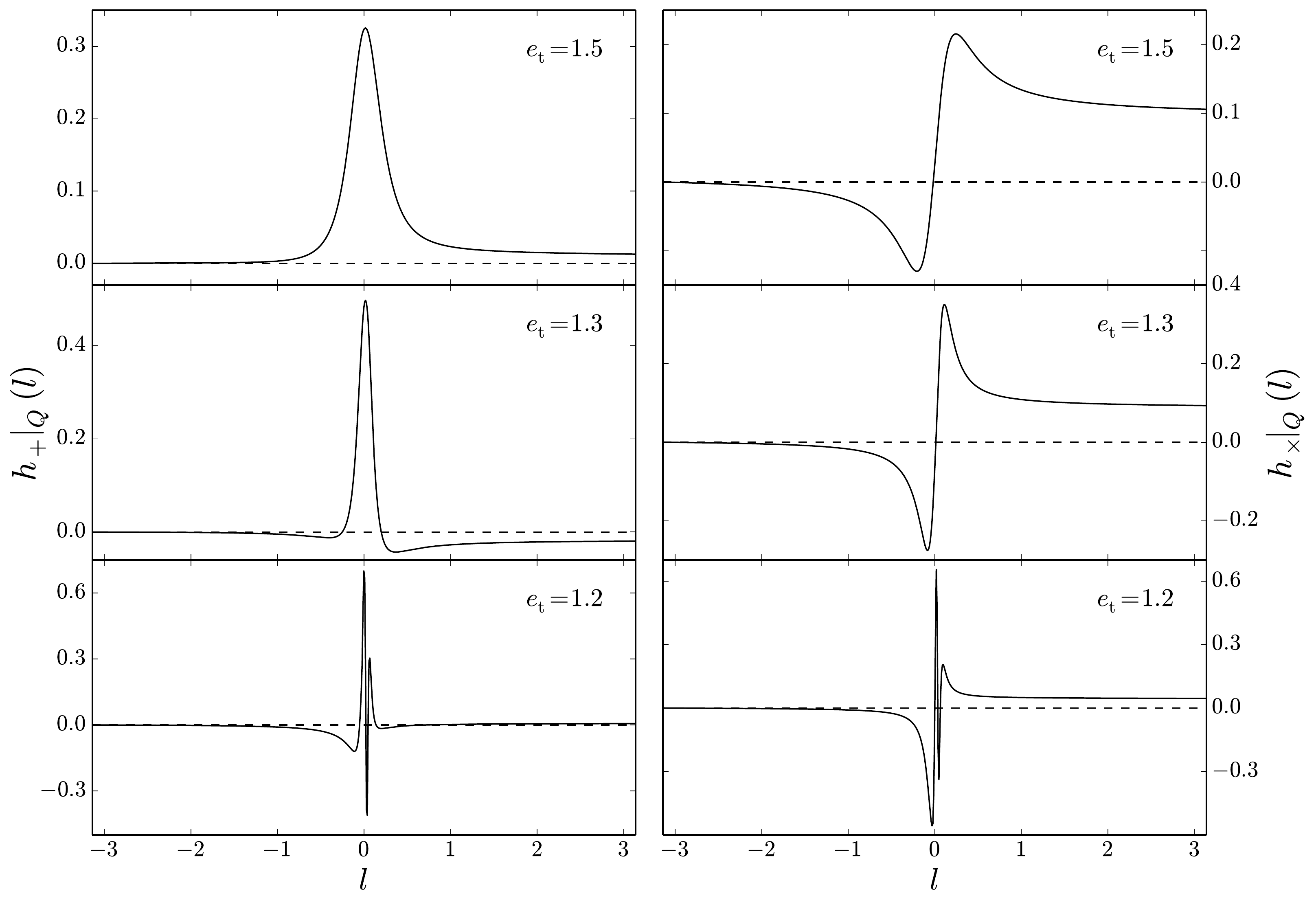}
    \caption{The scaled $h_+|_Q(l) $ and $h_{\times}|_Q(l) $ plots for $m= 20\, M_{\odot}, q=1$ spinning
      compact binaries containing maximally spinning BHs.
The employed scale factor in the present and the next two figures is $ G m /c^2 R$.
 We let $e_{\rm t} $ take three values while
      choosing $b \sim 30 \;Gm/c^2$. The initial spin orientations in the $\bm j_0$ based inertial frame are
      $\theta_1^i=30^{\circ}$, $\theta_2^i=30^{\circ}$, $\phi_1^i=30^{\circ}$, $\phi_2^i=120^{\circ}$ and we let $\theta = 45^{\circ}$.
      The conservative orbital evolution is fully 1.5PN accurate and the influences of GW emission are taken into account at leading order.
The  linear memory effect causes 
 the solid line waveform plots to depart from the dashed line after the hyperbolic passage.}
    \label{fig:sp-q1}
    \vspace{-0.2in}
  \end{figure}
  \begin{figure}[H]
    \centering
    \includegraphics[width=.77\textwidth]{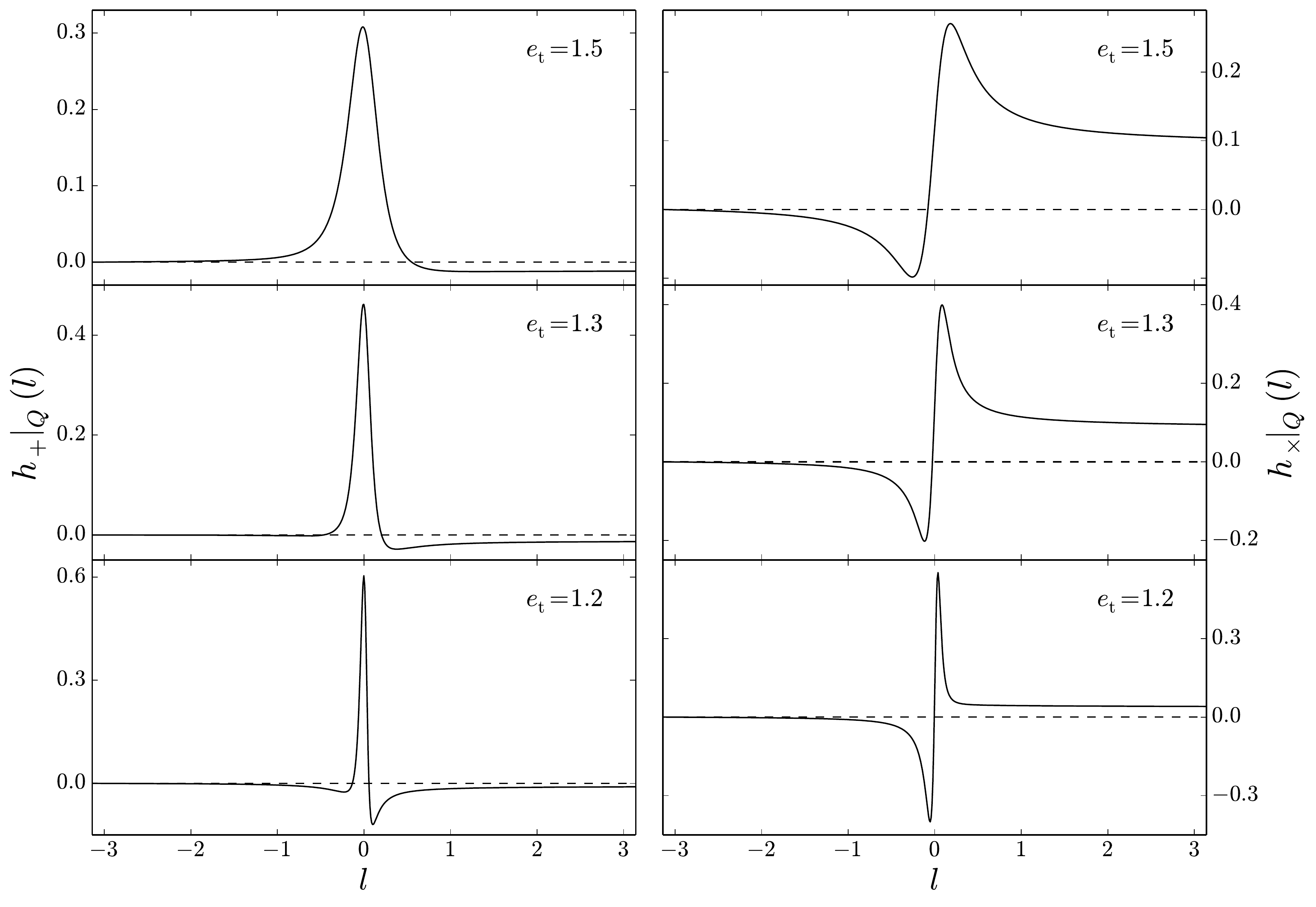}
    \caption{The scaled $h_+|_Q(l) $ and $h_{\times}|_Q(l) $ plots for $m= 20\, M_{\odot}, q=4$ spinning
	compact binaries containing maximally spinning BHs. The other specifications are same as in Fig.~\ref{fig:sp-q1}.
	The amplitude of the memory effect is rather insensitive to the mass ratio.
	However, the sharply varying features with multiple peaks, present in $e_t=1.2$ plots of Fig.~\ref{fig:sp-q1},
       are not visible.}
    \label{fig:sp-q4}
  \end{figure}
\twocolumngrid

\begin{figure}[h]
  \includegraphics[width=.5\textwidth]{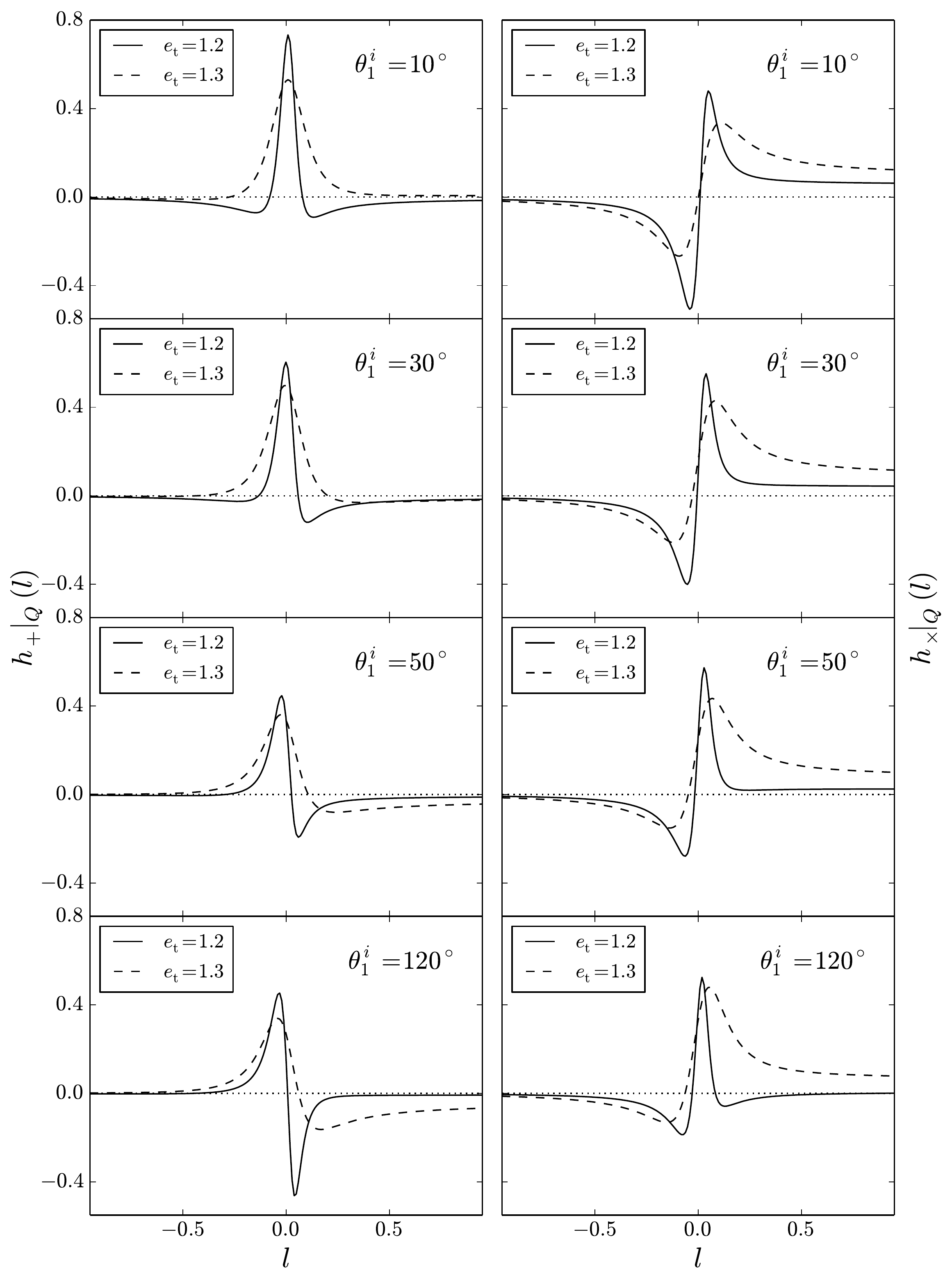}
    \caption{We plot the scaled $h_+|_Q(l) $ and $h_{\times}|_Q(l) $  for $m= 20\, M_{\odot}, q=4$ spinning
      compact binaries containing maximally spinning BHs while varying the initial orientation of the dominant spin for two $e_t$ values.
      The orbital dynamics is fully 1.5PN accurate while other parameters are similar to those used in Fig.~\ref{fig:sp-q4}.
      The impact of the initial misalignment between $\bm{s}_1$ and $\bm{j}_0$ on the memory 
      is more prominent on $h_+|_Q(l) $  for higher $e_t$ values.}
  \label{fig:spinningAngles}
\end{figure}

In Figs.~\ref{fig:sp-q1} and \ref{fig:sp-q4}, we display $h_+|_Q(l) $ and $h_{\times}|_Q(l) $ for 
$q=1$ and $q=4$ binaries having three different orbital eccentricities. We observe that 
both polarization states exhibit the memory effect. 
It is due to this `linear' memory effect that GW amplitudes at $t = + \infty$ are different from their respective 
$t = - \infty$ values, depicted by the dashed lines.
The amplitude of the effect decreases as the orbital eccentricity approaches unity. However, the time-domain waveforms 
develop sharply varying features compared to their higher orbital eccentricity counterparts.
This feature is mass ratio dependent and clearly visible for comparable mass binaries.
For a given $e_{\rm t}$ and $\bar n$, 
the amplitude of the memory effect is larger for $h_{\times}|_Q(l) $ compared to its `plus' counterpart.
Note that the memory effect is absent in $h_+|_Q(l) $ for non-spinning binaries
as evident from figures in Ref.~\cite{JS92,Favata11}.
In Fig.~\ref{fig:spinningAngles}, we probe the influence
of the initial orientation of the dominant spin on the memory effect 
for $q=4$ unequal mass binaries.
The memory amplitude decreases as we vary the initial misalignment of $\bm s_1$ from $\bm j_0$
for $e_{\rm t}$ values closer to unity.
The variations in the memory amplitude is more prominent for the $h_+|_Q(l) $ plots 
for higher $e_{\rm t}$ values. 
This may be attributed to the more pronounced orbital precession for higher 
$\theta_1^i$ values and the presence of non-negligible 
$\sqrt {e_{\rm t}^2 -1} $ contribution 
in the differential equations for the two spins, as evident from Eqs.~(\ref{eq:prec}).
It will be interesting to probe any possible data analysis implications of the varying memory 
amplitudes as depicted in Fig.~\ref{fig:spinningAngles}.
We have also verified that the memory effect persists even if we switch off the effects of GW damping.
In fact, the plots for $h_+|_Q(l) $ and $h_{\times}|_Q(l) $ are essentially identical 
to those displayed in Figs.~\ref{fig:sp-q1}, \ref{fig:sp-q4} and \ref{fig:spinningAngles} while 
neglecting the effects of GW damping, provided by Eqs.~(\ref{eq:radiation_reaction}).

In Appendix~A, we provide formulae for computing 1PN order amplitude corrected 
GW polarization states for spinning binaries in a compact way. 
The expressions, given by Eqs.~(\ref{eq:hxp1PN}), are obtained with the help of Eqs.~(\ref{eq:definition_hp_hx}) while using fully 1PN accurate 
expression for the transverse--traceless part of the radiation field  $h_{ij}^\text{TT}$.
The 1PN accurate expression for $h_{ij}^\text{TT}$ incorporates contributions from appropriate time derivatives of various
mass type and current type multipole moments of the binary and are adapted from Eqs.~(3.22) in Ref.~\cite{K95}.
Specifically, 1PN-accurate expression for $h_{ij}^\text{TT}$ requires us to compute  
time derivatives of mass and current quadrupoles and octopoles of the binary 
and fourth time derivative of  $\mathcal{I}^{ijkl}$ which is given by the symmetric and trace free 
part of $\mu \left (1 - 3\, \eta \right )\, x^{ijkl}$.
We do not provide 
the explicit 1PN accurate amplitude contributions to $h_+(l) $ and $h_{\times}(l) $ in terms of 
$\dot r , z, r\,\dot \Phi$ and trigonometric functions of $\iota, \alpha$ and $\Phi$ 
as done in Eqs.~(2.1) and (2.2). 
The very lengthy nature of such expressions is the main reason why we did not expand the squares and products
appearing in Eqs.~(\ref{eq:hxp1PN}) with the help of various dot products, given by Eqs.~(A2)-(A7). 
It is fairly straightforward to obtain plots for the 0.5PN and 1PN contributions to $h_+(l) $ and $h_{\times}(l) $,
similar to our plots for $h_+|_Q(l) $ and $h_{\times}|_Q(l) $. This is also not pursued as these PN contributions, as expected,
are substantially smaller in magnitude compared to the quadrupolar order waveforms.
Additionally, the plots for such PN contributions are qualitatively similar to plots for $h_+|_Q(l) $ and $h_{\times}|_Q(l) $
as these contributions also exhibit the linear memory effect.

In what follows, we provide an explanation for the presence of the linear memory effect in both the polarization states 
for spinning compact binaries in hyperbolic orbits.
This explanation becomes clearer 
and easier while considering non-spinning compact binaries in hyperbolic orbits.
In the next section, we consider temporally evolving GW polarization states associated with non-spinning compact 
binaries in 1PN accurate hyperbolic orbits invoking our Keplerian type parametric solution.
 This should also allow us to compare our
1PN accurate $h_+(l) $ and $h_{\times}(l) $ with those obtained via 
 the generalized true anomaly parameterization, detailed 
in Ref.~\cite{MFV10}.

\section{1PN accurate Gravitational wave phasing for non-spinning compact binaries in hyperbolic orbits}\label{sec:nonspinning}

  We begin by constructing quadrupolar order GW polarization states, namely $h_+|_Q(l) $ and $h_{\times}|_Q(l) $,
associated with non-spinning compact binaries moving in 1PN accurate hyperbolic orbits.
It is not very difficult to infer that Eqs.~(\ref{eq:h_plus_and_h_cross_in_n_and_v}) that provide $h_+|_Q(l) $ and $h_{\times}|_Q(l) $
in terms of various dot products involving $\bm{p}, \bm{q}, \bm{n}, \textbf{v}$ and $z$ also apply for 
non-spinning compact binaries.
Therefore, the explicit expressions for the quadrupolar order GW polarization states in terms of 
the relevant dynamical variables are
again obtained by evaluating the dot products appearing in Eqs.~(\ref{eq:h_plus_and_h_cross_in_n_and_v}).
It is natural to invoke an inertial frame associated with $\bm{L}$ to describe 
the orbital dynamics of non-spinning compact binaries. This is because $\bm{L}$ is conserved
both in magnitude and direction for non-spinning compact binaries. 
Furthermore, it 
is convenient to introduce
polar coordinates $(r, \phi)$ in a plane perpendicular to $\bm{L}$ as the orbital motion 
takes place in such a plane.
This allows us to describe $\bm{r}$ and $\textbf{v}$ in terms of $r, \phi$ and their time derivatives
in the $\bm{L}$ based inertial triad.
However, it is customary to evaluate the dot products
appearing in Eqs.~(\ref{eq:h_plus_and_h_cross_in_n_and_v})
 by expressing $\bm{r}$ and $\textbf{v}$
in an $\bm{N}$ based inertial frame $( \bm{p}, \bm{q}, \bm{N})$.
This is achieved by noting that the angle $\theta$  between $\bm{N}$ and $\bm{k}$, namely the orbital inclination,
remains a constant for non-spinning compact binaries.
This leads to the following expressions for  $\bm{r}$ and $\textbf{v}$ in the $( \bm{p}, \bm{q}, \bm{N})$ frame.
\begin{subequations}
  \begin{align}
    \bm{n} =& \;-\bm{p}\sin\phi + (\bm{q}\cos \theta + \bm{N}\sin \theta) \cos\phi~, \\
    \textbf{v} =&\; \bm{p}\left(-\dot{r}\sin\phi-r\dot{\phi}\cos{\phi}\right)\\
		 &+\,\left(\bm{q}\cos\theta+\bm{N}\sin\theta \right) \left(\dot{r}\cos\phi-r\dot{\phi}\sin\phi\right)~.\nonumber
  \end{align}
\end{subequations}

The above expressions allow us to
compute the dot products appearing in the Eqs.~(\ref{eq:h_plus_and_h_cross_in_n_and_v}) for $h_+|_Q(l) $ and $h_{\times}|_Q(l) $,
in a straightforward manner.
The resulting GW polarization states read
\begin{subequations}
  \label{11}
  \begin{align}
    h_+|_{\rm Q} = \;&\frac{G\mu}{c^4R}\;\times  \label{9}\\
		   &\bigg\{\left(1+C_{\theta}^2\right)\bigg[\left(z+r^2\dot{\phi}^2-\dot{r}^2\right)\cos2 \phi\;\nonumber \\
		   &+2r\dot{r}\dot{\phi}\sin 2\phi\bigg] - S_{\theta}^2\left(z-r^2\dot{\phi}^2-\dot{r}^2\right)\bigg\}~, \nonumber
  \end{align}
  \begin{align}
    h_{\times}|_{\rm Q} = \;&2\frac{G\mu}{c^4R}C_{\theta}\;\times   \label{8} \\
			  &\bigg\{\left(z+r^2\dot{\phi^2}-\dot{r}^2\right)\sin 2\phi-2r\dot{r}\dot{\phi}\cos 2\phi\bigg\}~,\nonumber
  \end{align}
\end{subequations}
where $\dot{\phi} = \text{d}\phi/\text{d}t$.

  It should be obvious that we need to describe how $r, \dot r, \phi$ and $\dot \phi$ 
evolve in time for non-spinning compact binaries 
moving in  hyperbolic orbits to 
obtain the associated $h_+|_Q(t) $ and $h_{\times}|_Q(t) $. 
This is implemented in a parametric manner invoking 
the 1PN accurate quasi-Keplerian parameterization of Ref.~\cite{DD85}.
The radial and angular parts of the orbital dynamics are parametrically given by 
\begin{subequations}
  \label{eq:radang_ns}
  \begin{align}
    r =& \;a_{\rm r}(e_{\rm r}\cosh v-1)~, \\
    \phi-\phi_0 =& \; 2K\arctan\biggl[\left(\frac{e_{\phi}+1}{e_{\phi}-1}\right)^{1/2}\tanh v/2\biggl]~.
  \end{align}
\end{subequations}
The temporal evolution for $r$ and $\phi$ are  provided numerically by tackling the 1PN accurate Kepler equation
\begin{equation}
  l = \bar n(t-t_0)=e_{\rm t}\sinh v-v. \label{33}
\end{equation}
The additional orbital parameters $K $ and $e_{\phi}$ that appear in the angular part of the parametric 
solution are the hyperbolic versions of the periastron advance constant and angular eccentricity 
associated with the eccentric orbits \cite{DD85}.
All the orbital elements, as expected, are PN accurate functions of conserved orbital energy and angular momentum.
The 1PN accurate expressions for these orbital elements in terms of the conserved energy and 
angular momentum are provided by Eqs.~(3.6) and (4.13) in Ref.~\cite{DD85}.
These inputs allow us to compute  1PN accurate expressions for 
$r, \dot r, \phi$ and $\dot \phi$ in terms of $v, e_{\rm t}, \bar n, m$ and $\eta$. 
The explicit expressions of these dynamical variables are 
\onecolumngrid
\begin{subequations}
  \label{eq:quasi_kepl_nonsp}
  \begin{align}
    r (v) =\; &\frac{Gm}{c^2}\frac{1}{\bar \xi^{2/3}}(e_{\rm t}\cosh v-1)\,\left\{1+\bar \xi^{2/3}\;\frac{2\eta-18-(6-7\eta)e_{\rm t}\cosh v}{6\left(e_{\rm t}\cosh v-1\right)}\right\}~, \label{1}\\
    \dot{r} (v) =\; &\bar \xi^{1/3}\frac{c\, e_{\rm t}\sinh v}{e_{\rm t}\cosh v-1}\left\{1-\bar \xi^{2/3}\frac{6-7\eta}{6}\right\}~,\\
    \phi (v) - \phi_0 =\; &2\arctan \left[\left(\frac{e_{\phi}+1}{e_{\phi}-1}\right)^{1/2}\tanh v/2\right]\,\left\{1+\bar \xi^{2/3}\frac{3}{e_{\rm t}^2-1}\right\}~,\\ 
    \dot{\phi} (v) =\; &\frac{\bar n\, \sqrt{e_{\rm t}^2-1}}{\left(e_{\rm t}\cosh v-1\right)^2}\bigg\{1-\bar \xi^{2/3}\,\frac{\left[3-\left(4-\eta\right)e_{\rm t}^2+\left(1-\eta\right)e_{\rm t}\cosh v\right]}{\left(e_{\rm t}^2-1\right)\left(e_{\rm t}\cosh v-1\right)}\bigg\}\label{4}~. 
  \end{align}
\end{subequations}
\twocolumngrid
To obtain Eqs.~(\ref{eq:quasi_kepl_nonsp}), we have used PN accurate relations connecting $e_{\rm r}$ and $e_{\phi}$ to $e_{\rm t}$, namely 
$e_{\rm r} = e_{\rm t} \{1-\bar \xi^{2/3}(8-3\eta)/2\} $ and $e_{\phi} = e_{\rm t} \{1-\bar \xi^{2/3}(4-\eta)\}$.
The fact that we have invoked $e_{\rm t}$ to characterize the dynamics allows us to invoke Mikkola's approach to 
numerically solve the classical Kepler equation for hyperbolic orbits as detailed in Sec. 4 in Ref.~\cite{Mikkola}.
We use the resulting $v(l)$ in 
Eqs.~(\ref{eq:quasi_kepl_nonsp})
 to obtain 1PN accurate $l$ evolution for 
$r, \dot r, \phi$ and $\dot \phi$ for a non-spinning compact binary characterized by $m, \eta, \bar n$ and $e_{\rm t}$.
These evolutions are implemented in Eqs.~(\ref{11}) to obtain  $h_+|_Q(t) $ and $h_{\times}|_Q(t) $ associated
with compact binaries moving in 1PN accurate hyperbolic orbits.

   We move on to compute explicit expressions for $h_+ $ and $h_{\times}$ that are also 1PN accurate in their amplitudes.
This requires us to implement Eqs.~(\ref{eq:definition_hp_hx}) for $h_+ $ and $h_{\times} $ while using the 1PN accurate expression for 
$h_{ij}^\text{TT}$ for general orbits that are available in Refs.~\cite{WW96,GI97}.
The resulting 1PN accurate amplitude corrected expressions for $h_+ $ and $h_{\times}$ can be written as 
\begin{subequations}\label{eq:polstates_exp}
  \begin{align}
    h_+  &= \frac{G\mu}{c^4R}\, \left ( h_+^N + \frac{1}{c}h_+^{0.5} + \frac{1}{c^2}h_+^{1} \right )~,\\
    h_{\times}  &= \frac{G\mu}{c^4R}\, \left ( h_{\times}^N + \frac{1}{c} h_{\times}^{0.5} + \frac{1}{c^2} h_{\times}^{1} \right )~,
  \end{align}
\end{subequations}
where $ h_{+,\times}^N, h_{+,\times}^{0.5}$ and $ h_{+,\times}^{1}$ are given by
\onecolumngrid
\begin{subequations}
\begin{align}
  h_+^N &=  2 r\dot{r}\dot{\phi} (1+C_{\theta}^2)\sin{2\phi}+(1+C_{\theta}^2)\left(z+r^2\dot{\phi}^2-\dot{r}^2\right)\cos{2\phi}+S_{\theta}^2\left(\dot{r}^2+r^2\dot{\phi}^2-z\right)~,\\[0.2in]
  h_{\times}^N &= 2C_{\theta}\left(z+r^2\dot{\phi}^2-\dot{r}^2\right)\sin{2\phi}-2C_{\theta}2r\dot{r} \dot{\phi} \cos{2\phi}~,
\end{align}
\begin{align}
  h_+^{0.5PN} &= \frac{S_{\theta}}{2} (X_1-X_2) \bigg[(3 C_{\theta}^2-1) \left(\dot{r}^2+r^2\dot{\phi}^2-2z\right) \dot{r} \cos{\phi}+(1+C_{\theta}^2) \left(\dot{r}^2-3r^2\dot{\phi}^2-2z\right) \dot{r} \cos{3\phi}\label{eq:hpns05PN}\\
 &- \left((\dot{r}^2+r^2\dot{\phi}^2-z)(3C_{\theta}^2-1)-z(C_{\theta}^2+5)\frac12\right)r\dot{\phi}\sin{\phi}+(1+C_{\theta}^2)\left(3\dot{r}^2-r^2\dot{\phi}^2-\frac72z\right)r\dot{\phi}\sin{3\phi}\bigg]~,\nonumber\\[0.2in]
  h_{\times}^{0.5PN} &= \frac{C_{\theta} S_{\theta}}{2} (X_1-X_2) \bigg[\left(2\dot{r}^2+2r^2\dot{\phi}^2-5z\right) r\dot{\phi}\cos{\phi}+\left(6\dot{r}^2-2r^2\dot{\phi}^2-7z\right)r\dot{\phi}\cos{3\phi}\label{eq:hxns05PN}\\
       &+2\left(\dot{r}^2+r^2\dot{\phi}^2-2z\right)\dot{r}\sin{\phi}+2\left(\dot{r}^2-3r^2\dot{\phi}^2-2z\right)\dot{r}\sin{3\phi}\bigg]~,\nonumber
\end{align}
\begin{align}
  h_+^{1PN} &= \frac{1}{24} \bigg[-18 S_{\theta}^4 (\dot{r}^4+r^4\dot{\phi}^4)(3\eta-1)+r^2\dot{\phi}^2 z S_{\theta}^2 (51\eta-69+C_{\theta}^2 39(1-3\eta))+z^2\times\label{eq:hpns1PN}\\
     & S_{\theta}^2(116+7(1-3\eta)(1-3C_{\theta}^2))-36 \dot{r}^2 S_{\theta}^4 r^2\dot{\phi}^2(3 \eta-1)+6\dot{r}^2 z S_{\theta}^2(9C_{\theta}^2(1-3\eta)+3+13\eta)\bigg]\nonumber\\
     &+\cos{2\phi}\;\frac12\bigg[(\dot{r}^4-r^4\dot{\phi}^4)(1+C_{\theta}^2(1+2S_{\theta}^2))(3\eta-1)+r^2\dot{\phi}^2z\bigg((2+3\eta)(1+C_{\theta}^2)+S_{\theta}^2(3\eta-1)(4+C_{\theta}^2)\bigg)\nonumber\\
     &-\dot{r}^2z\bigg((3+2\eta)(1+C_{\theta}^2)+6C_{\theta}^2S_{\theta}^2(3\eta-1)\bigg)+z^2\frac13\bigg(7C_{\theta}^2S_{\theta}^2(3\eta-1)-29(1+C_{\theta}^2)\bigg)\bigg]\nonumber\\
     &+\cos{4\phi}\;\frac{(1+C_{\theta}^2)S_{\theta}^2}{24}(3\eta-1)\bigg[6(\dot{r}^4+r^4\dot{\phi}^4)+51r^2\dot{\phi}^2z+7z^2-18\dot{r}^2(2r^2\dot{\phi}^2+z)\bigg]\nonumber\\
     &-\sin{2\phi}\;r\dot{r}\dot{\phi}\bigg[(1+C_{\theta}^2(1+2 S_{\theta}^2))(\dot{r}^2+r^2\dot{\phi}^2)(3\eta-1)-z\bigg((2+4\eta)(1+C_{\theta}^2)+\frac{S_{\theta}^2}{2}(3\eta-1)(1+9C_{\theta}^2)\bigg)\bigg] \nonumber\\
     &-\sin{4\phi}\;r\dot{r}\dot{\phi}\,(3\eta-1)(1+C_{\theta}^2)S_{\theta}^2\left(\dot{r}^2-r^2\dot{\phi}^2-\frac94z\right)~,\nonumber\\[0.5in]
  h_{\times}^{1PN} &= zr\dot{r}\dot{\phi}(1-3\eta)\frac{C_{\theta}S_{\theta}^2}{2}+\cos{2\phi}\;r\dot{r}\dot{\phi}\,C_{\theta}\bigg[2(1+S_{\theta}^2)(3\eta-1)(\dot{r}^2+r^2\dot{\phi}^2)-z\bigg(4+8\eta+5S_{\theta}^2(3\eta-1)\bigg)\bigg]\label{eq:hxns1PN}\\
     &+\cos{4\phi}\;r\dot{r}\dot{\phi}(3\eta-1)C_{\theta}S_{\theta}^2\left[2\dot{r}^2-2r^2\dot{\phi}^2-\frac92z\right]+\sin{2\phi}\;C_{\theta}\bigg[(\dot{r}^4-r^4\dot{\phi}^4)(1+S_{\theta}^2)(3\eta-1)+z\frac{r^2\dot{\phi}^2}{2}\times\nonumber\\
     &(4+6\eta+5S_{\theta}^2(3\eta-1))+z^2\frac16(-58+7S_{\theta}^2 (3\eta-1))-\dot{r}^2z(3+2\eta+S_{\theta}^2 (9\eta-3))\bigg]\nonumber\\
     &+\sin{4\phi}\;(3\eta-1)\frac{C_{\theta}S_{\theta}^2}{12}\bigg[6\dot{r}^4+6r^4\dot{\phi}^4+51r^2\dot{\phi}^2z+7z^2-18\dot{r}^2\bigg(2r^2\dot{\phi}^2+z\bigg)\bigg]~.\nonumber
\end{align}
\end{subequations}
\twocolumngrid
We have verified that in the circular limit the above expressions reduce to Eqs.~(3) and (4) in Ref.~\cite{BIWW96}.
This requires us to equate $ \dot r $ and  $ \dot \phi $ to zero and  $ v/r $, respectively, while replacing
$\phi$ by $\phi + \pi/2$. This is done to make sure that the orbital phase is measured from the
same axis as in \cite{BIWW96}. Afterwards, we need to connect $ v$ and $ z $ by the 1PN-accurate relation 
$ v = z^{1/2} + z^{3/2}\, \left ( \eta - 3 \right )/( 2\, c^2) $ and express $ z $ to the variable
$ x= ( G\, m\, \dot \phi/c^3 )^{2/3} $ of Ref.~\cite{BIWW96} through
the 1PN-accurate relation $ z = c^2\, x \left (  1 + ( 3- \eta )\,x/3 \right ) $.\\

\begin{figure}[H]
    \includegraphics[width=\textwidth]{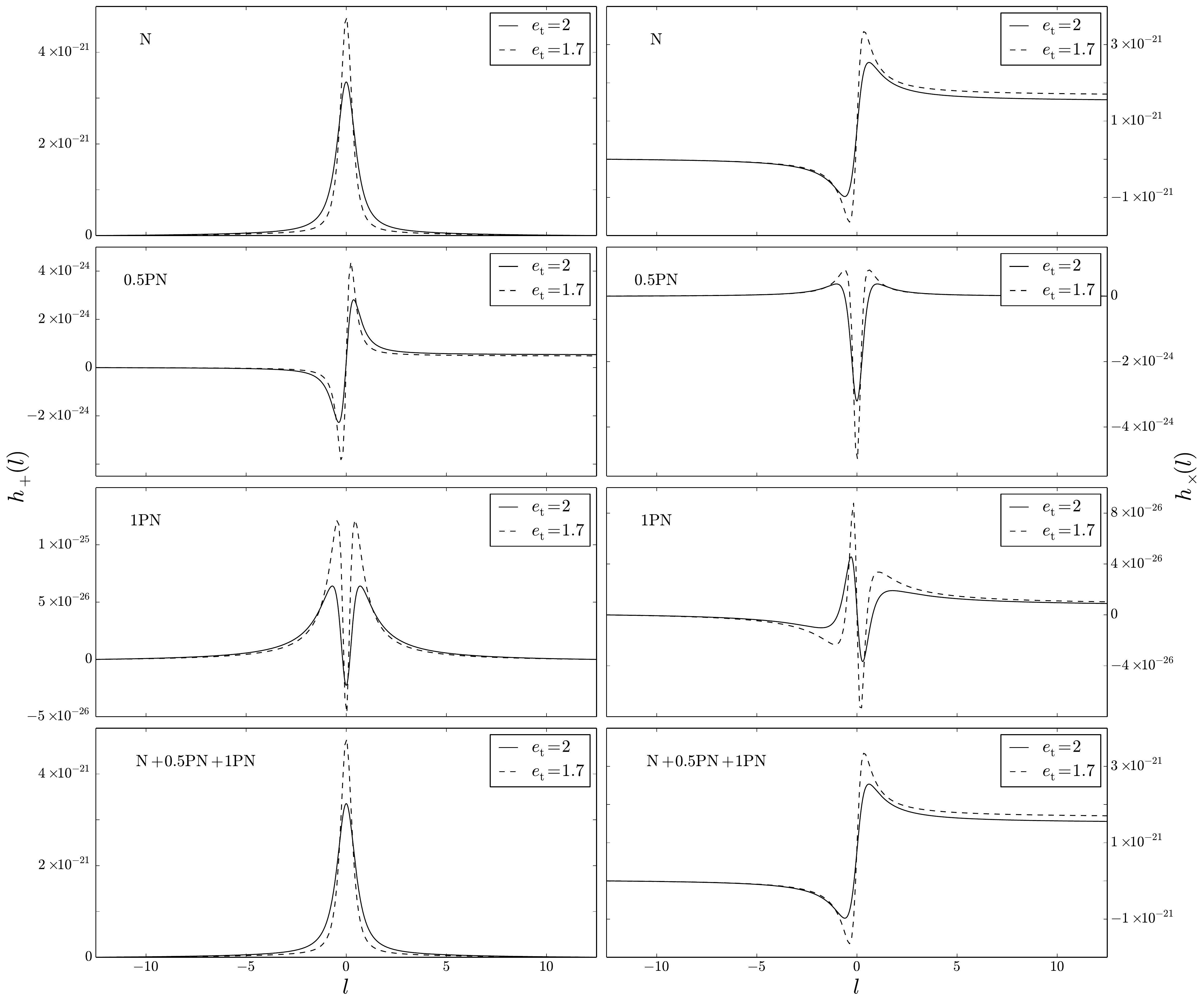}
    \caption{Polarization states at Newtonian, 0.5PN and 1PN order, as well as their sum, respectively, as functions of $l$ for non-spinning compact binaries
    in hyperbolic orbits. The solid line shows the case where $e_{\rm t} = 2$ and the dashed line 
    shows the waveform for $e_{\rm t} = 1.7$. The masses are $m_1 = 8\,M_{\odot}$ and $m_2 = 13\,M_{\odot}$, while 
    the minimal distance is chosen to be $r_{\text{min}} \sim 1.9\times 10^9 m$, as in Ref.~\cite{MFV10}.
    The waveform for the $h_+$ polarization is shown on the left and $h_{\times}$ appears on the right.}
    \label{fig:ns}
  \end{figure}

We are now in a position to plot the Newtonian, 0.5PN and 1PN contributions to GW polarization states for non-spinning
compact binaries moving in 1PN accurate hyperbolic orbits. This is pursued in Fig.~\ref{fig:ns} for a binary having 
$m_1= 8 M_{\odot}, m_2 = 13 M_{\odot}$ to compare with Figs.~6-10 in Ref.~\cite{MFV10} 
while choosing  $e_{\rm t}$ to be $1.3$ and $2$.
The first three rows display the Newtonian, 0.5PN and 1PN contributions to $h_+ $ and $h_{\times}$
for binaries having fully 1PN accurate orbital evolution while the amplitude contributions are 
fully 1PN accurate for the fourth row plots.
Apart from the change in their amplitudes, there are no changes in the way various contributions temporally 
evolve as we vary the orbital eccentricity. 
To make sure of the correctness of our approach, we have reproduced temporal
evolution for the real and imaginary parts of 
the time derivatives of mass and current multipole moments
that are displayed in Fig.~8 of \cite{JS92}. Additionally, we are also able to reproduce 
temporal evolution for 
the real and imaginary parts of the $(2,2)$ GW mode depicted in  Fig.~2 of \cite{Favata11} by our approach.\\

However, a visual comparison of our $e_{\rm t}=2$ plots that appear in the first three rows of Fig.~\ref{fig:ns} 
with similar plots in Figs.~6,7 and 10  of Ref.~\cite{MFV10} reveals substantial differences.
The differences are clearly noticeable for the cross polarization states.
Interestingly, plots in Figs.~8 of Ref.~\cite{MFV10} that display what they describe as the multipolar 
1PN corrections to GW polarization states are qualitatively in agreement with $e_{\rm t}=2$ plots in the third row 
of our Fig.~\ref{fig:ns}. The nature of the memory effect exhibited by the Newtonian contribution to $h_{\times}$,
as shown in Fig.~6 of Ref.~\cite{MFV10},
is also qualitatively different from our plots and those available in the literature.
We suspect that the observed differences may be due to the way temporal evolution is implemented in Ref.~\cite{MFV10}.
Note that this is implemented analytically as a PN series 
in terms of the coordinate time 
as evident from the PN accurate expression for their angular variable \cite{MFV10}.
However, we describe orbital dynamics in a parametric
way and invoke numerical solution of the PN accurate Kepler equation to obtain the time evolution.
It will be interesting to probe why these approaches differ for hyperbolic orbits.

\onecolumngrid

\twocolumngrid
We turn to explain the presence of \emph{linear memory effect} 
in both the quadrupolar order polarization states of spinning compact binaries in hyperbolic orbits.
We begin by explaining, in detail, why only the cross polarization state 
exhibits the effect in the case of non-spinning compact binaries during hyperbolic
encounters with the help of Refs.~\cite{Favata09, Favata11}.
This diversion is desirable as we can pinpoint the terms that explicitly cause
the memory effect for such binaries.
Unfortunately, this is rather impossible in the case of  spinning compact binaries in hyperbolic orbits
as several dynamical variables, present in the Eqs.~(2.1) and 
(2.2) for $h_+|_{\rm Q}$ and  $h_{\times}|_{\rm Q}$,
can contribute to this effect.
We begin by noting that an ideal GW detector will return to its 
original configuration after the passage of an incident GW, 
if the signal does not exhibit any memory effect.
In contrast, transient GW signals that possess memory effects
force the detector not to relapse to the initial  
configuration even after the passage of the GW.
This is essentially due to a net change in the amplitude of the local 
metric induced by the passage of such a GW.
This leads to the linear memory effect that we observe in Fig.~\ref{fig:ns}.

  Influenced by Ref.~\cite{Favata09} and  with the help of Eq.~(2.4), we write the net change 
in the quadrupolar order far-zone radiation field as 
\begin{equation}
  \Delta h_{ij}^{TT}|_Q =  \frac{G}{c^4}\frac2R \Delta(\mathcal{\ddot{I}}_{ij}^{TT})~,
\end{equation}
where $\mathcal{I}_{ij}$ is the mass-quadrupole moment of the binary, given by
$\mathcal{I}_{ij} = \mu\, x_i\, x_j$ at the  Newtonian order.
It is fairly straightforward to compute the second time derivative 
of the transverse-traceless part of $\mathcal{I}_{ij}$ 
using Newtonian equations of motion: $\ddot{x}_i= -G\,m\,x_i/r^3$.
The resulting expression reads 
\begin{equation}\label{eq:Iij_derivative}
  \ddot{\mathcal{I}}^{TT}_{ij} = 2 \mu \big(\dot{x}_i\,\dot{x}_j - \frac{G\,m}{r^3}x_i\,x_j\big)
\end{equation}
The second term in Eq.~(3.9) vanishes for $t\to\pm\infty$
as it falls off like $1/r$.
This is because $v \to\pm\infty $ as $t\to\pm\infty$ due to Eq.~(3.4) and 
$r (v)$ is proportional to $(e_t\, \cosh v - 1)$ as evident from Eqs.~(\ref{eq:quasi_kepl_nonsp}).
However,  the
magnitude of the 
 relative velocity approaches
a finite value, namely $\text{v}_{\infty}=\sqrt{2E}$ 
(note that $E$ stands for the orbital energy scaled by $\mu$).
This results in the following expression for the Newtonian order linear memory
effect associated with hyperbolic passages
\begin{equation}\label{eq:general_memory}
  \Delta h_{ij} = 4\frac{G \mu}{c^4 R} \Delta(\dot{x}_i\,\dot{x}_j)\,.
\end{equation}
Clearly, the differences in the components of the orbital velocity 
as $t\to\pm\infty$ contribute substantially to the magnitude of memory effect.

 To demonstrate explicitly the memory effect, let us consider the following scenario
where the observer is perpendicular to the orbital plane ($\theta =0$ orientation).
We infer that the non-zero components of $\mathcal{I}_{ij}$ are 
 $\mathcal{I}_{11}, \mathcal{I}_{22}$ and $\mathcal{I}_{12}=\mathcal{I}_{21}$,
where indices $1$ and $2$ stand for the $x$ and $y$ components in the
$\bm j_0$-based inertial frame.
Additionally, $x$ and $y$ components of the orbital velocity as 
$t\to\pm\infty$ are given by
\begin{subequations}
\begin{align}
  \lim_{t\to\pm\infty} \dot{x}_1 &= \pm \, \text{v}_{\infty}\,\cos{\pm \phi_{\infty}}~,\\
  \lim_{t\to\pm\infty} \dot{x}_2 &= \pm \, \text{v}_{\infty}\,\sin{\pm \phi_{\infty}}~\,,
\end{align}
\end{subequations}
where $\phi_{\infty}$ stands for the asymptotic value for the orbital phase
as $t\to\pm\infty$, which can be deduced from eqs.~(\ref{eq:radang_ns}) and (3.4).
With these inputs and some trigonometric manipulations, we 
obtain the following expressions for the linear memory amplitudes associated with 
different components of $h_{ij}$ 
\begin{subequations}\label{eq:memory_favata}
\begin{align}
  \Delta h_{ii} &= 0~, &i&=1,2~,\\
  \Delta h_{ij} &= -\,8\frac{G\, \mu }{c^4}\frac{E}{R}\sin{2\phi_{\infty}}~, &i&\neq j~.
\end{align}
\end{subequations}
In what follows we show that our expressions for the quadrupolar order 
$h_+$ and  $h_{\times}$, given by Eqs.~(\ref{11}), are indeed consistent with 
the above estimates.

 To make contact with the above discussions, we consider again the binary 
configuration having  $\theta=0$.
It is fairly straightforward to obtain $t\to \pm \infty$ 
limits of our quadrupolar 
order expressions for $h_+|_Q(t) $ and $h_{\times}|_Q(t ) $, given by Eqs.~(\ref{11}).
It may be recalled that $v$ also approaches $\pm \infty$ as $t\to \pm \infty$ due to 
Eq.~(3.4). This ensures that 
the dynamical variables $z (v) = G\,m/r (v), \dot{\phi}(v)$ and
 the product $r(v)\dot{\phi}(v)$, displayed in Eqs.~(\ref{eq:quasi_kepl_nonsp}),
 go to zero in the limit $ v \to \pm \infty$.
This is of course due to the presence of $(e_t\, \cosh v -1)$ and its powers in the 
denominators of these parametric expressions.
However, the expression for $\dot{r}(v)$ does not vanish 
as $ t \rightarrow \pm \infty$, but rather tends to the finite value, namely $\text{v}_{\infty}$.
This forces the expressions for  $h_+|_Q(t) $ and $h_{\times}|_Q(t) $, given by 
Eqs.~(\ref{11}), at $t\to +\infty$ to be 
\begin{subequations}
\begin{align}
  h_+|_Q &= - 2\frac{G\mu}{c^4R} \text{v}_{\infty}^2\, \cos{2\phi_{\infty}} \label{eq:hp_inf}~,\\
  h_{\times}|_Q  &= - 2\frac{G\mu}{c^4R} \text{v}^2_{\infty}\, \sin{2\phi_{\infty}}~.\label{eq:hc_inf}
\end{align}
\end{subequations}

Note that the right hand side of Eq.~(3.13a) is an even function of $\phi$ 
due to the presence of $\cos{2\phi }$. However, the right hand side
of Eq.~(3.13b) is an odd function of $\phi$.
This implies that 
\begin{subequations}
\begin{align}
  \lim_{t\to +\infty} h_+|_Q(t) &= \lim_{t\to -\infty} h_+|_Q(t)~,\\
  \lim_{t\to +\infty} h_{\times}|_Q(t) &= - \lim_{t\to -\infty}h_{\times}|_Q(t)~\,.
\end{align}
\end{subequations}
Therefore, the amplitude differences in 
the above two polarization states between 
the early and late times during hyperbolic encounters are
\begin{subequations}\label{eq:memory_DGGJ}
  \begin{align}
    \Delta h_+ &= 0~,\\
    \Delta h_{\times} &= -\,8\frac{G\, \mu }{c^4}\frac{E}{R}\sin{2\phi_{\infty}}~,
  \end{align}
\end{subequations}
where we used the relation $\text{v}^2_{\infty}=2E$.
Clearly, the above two expressions are identical to Eqs.~(\ref{eq:memory_favata}) that 
we derived using the detailed discussions of Ref.~\cite{Favata09}.
This also explains the linear memory effect exhibited by the first row plots in Fig.~\ref{fig:ns}.

 It is possible to employ similar arguments to show that 0.5PN contributions to 
 $h_{+}$, given by Eq.~(3.7c),
 and 1PN contributions to $h_{\times}$, given by Eq.~(3.7f),
should exhibit the linear memory effects during the hyperbolic encounters.
This is essentially due to the presence of non-vanishing odd functions
$\dot {r}^3\, \cos \phi $ and $\dot {r}^3 \cos{3\,\phi}$, as well as
$\dot {r}^4\, \sin {2\,\phi}$ and $\dot {r}^4\, \sin {4\,\phi}$
in the above expressions.
Terms like $\dot{r}^3 \sin\phi$ or $\dot{r}^3 \sin{3\,\phi}$ appearing in the
expression for $h_{\times}$ at 0.5PN order will not
contribute to the memory, since both the $\dot{r}^3$ and the $\sin\phi$ factors
are odd functions of time and yield therefore an even term.
The second and third row plots in Fig.~\ref{fig:ns} clearly support the above inference.
In contrast, both $0.5$PN order GW polarization states, depicted in Fig.~7 in \cite{MFV10},
show the memory effect as evident from their dashed line plots.
This is also applicable to 1PN order corrections to $h_{\times}$ and $h_{+}$, as
displayed by the dashed line plots in Fig.~9 in \cite{MFV10}, that arise 
from perturbative description of their orbital elements.
Clearly, such plots are inconsistent with plots in our Fig.~\ref{fig:ns}.

 When we include the spin effects, 
it is rather difficult to obtain similar analytic estimates to demonstrate
why both polarization states should exhibit the linear memory effect.
However, note that the expressions for both 
$h_+|_Q(t) $ and $h_{\times}|_Q(t) $ contain terms like $\dot{r}^2\sin2\Phi(t)$
as evident from our Eqs.~(2.1) and (2.2).
Additionally, the phasing angle $\Phi(t)$ does not have the same value at
$ t = \pm \infty $ anymore due to the spin-orbit
coupling induced precession of the orbital plane.
These effects force the plots of $h_+|_Q(t) $ and $h_{\times}|_Q(t) $ to exhibit 
the linear memory effect as displayed in the previous section.

Finally, let us comment about the  impact of orbital eccentricity on 
the amplitude of the memory effect.
This should be easily extractable with the help of Eq.~(\ref{eq:memory_DGGJ})
and the fact that the angle at infinite times $\phi_{\infty}$
is related to $e_{\rm t}$ through the relation
$\cos\phi_{\infty} = -1/e_{\rm t}$. This relation is also 
equivalent to $\sin{2\phi_{\infty}}=-2\sqrt{e_{\rm t}^2-1}/e_{\rm t}^2$.
Therefore, the memory goes to $0$ when $e_{\rm t} \to 1$ and we have
$\Delta h_{\times} \propto 1/e_{\rm t}$ for $e_{\rm t} \gg 1$.
Moreover, the amplitude of the effect peaks at eccentricity $e_{\rm t} = \sqrt{2}$
and this is consistent with our results 
of the present and previous sections.
Let us also comment about the plausibility of observing  the
influences of memory effect.
Unfortunately, laser interferometric GW observatories are not the
ideal instruments to probe the implications of both linear and
non-liner memory effects as explained in Ref.~\cite{Favata09}.
This is essentially because the internal forces present in such instruments
are expected to bring the test masses back to their  original (or initial)
configurations. However, it may be possible to detect the implications
of non-linear memory effects associated with the merger of supermassive
black hole binaries with the help of the ongoing and planned pulsar
timing arrays (PTAs) \cite{Seto09}.
This is despite the fact that characteristic merger frequencies of such
binaries are far higher than the nano-Hertz regime, relevant for PTAs.

\section{Conclusions}\label{sec:conclusions}

We provided an efficient prescription to compute  GW polarizations 
that are PN accurate both in amplitude and phase evolution 
for spinning compact binaries in hyperbolic orbits. 
The incorporated spin effects are due to the dominant order spin-orbit interactions
while the non-spinning orbital dynamics is 1PN accurate.
The radial part of the conservative 1.5PN accurate orbital dynamics is treated in a parametric 
way by adapting the Keplerian type parametric solution for eccentric orbits, available in Ref.~\cite{GS11}.
We invoked Mikkola's accurate and efficient method to numerically solve the hyperbolic version of the 
Kepler Equation to obtain temporal evolution for $r$ and $\dot r$.
In contrast, the 1.5PN accurate angular sector of the dynamics is tackled numerically
by solving differential equations that describe the orbital phase evolution and 
the precessional dynamics of $\bm s_1, \bm s_2$ and $\bm k$.
We also incorporated the influence of GW emission on this 1.5PN accurate orbital dynamics. 
Afterwards, we numerically inserted the variables that describe the radial, angular and precessional
aspects of the orbital dynamics into 
PN accurate expressions for the two GW polarization states for compact binaries in general orbits.
This is how we constructed ready-to-use waveforms for spinning compact binaries in hyperbolic orbits.
    
We observed the presence of linear GW memory effect in both the polarization states. In contrast, only the cross
polarization state exhibits the memory effect for non-spinning compact binaries in hyperbolic orbits and 
we provided an explanation for these observations.
We explored the influence of orbital eccentricity, mass ratio and dominant spin orientation on the evolution of the two 
polarization states and the amplitude of the memory effect. 
Invoking the non-spinning version of our approach, we have 
reproduced the temporal
evolution for the real and imaginary parts of 
the time derivatives of mass and current multipole moments
and associated GW modes, 
detailed in Refs.~\cite{JS92,Favata11}.
However, various temporally evolving PN contributions to $h_+$ and $h_{\times}$
associated with non-spinning compact binaries, displayed in 
Figs.~6, 7, 9 and 10 of Ref.~\cite{MFV10},
are visually different from what we obtained. 
We provided a possible qualitative explanation for these differences.\\

It will be interesting to incorporate the 2PN order non-spinning
contributions to our 1.5PN accurate
orbital dynamics. This is rather tricky due to the appearance of 2PN order
$ f-u \equiv 2\tan^{-1}\left(\beta_{\phi}\sin u/(1-\beta_{\phi}\cos
u)\right)$ term in the
PN accurate Kepler Equation for eccentric binaries, where
$\beta_{\phi}=(1-\sqrt{1-e_{\phi}^2})/e_{\phi}$
and $f$ is the true anomaly \cite{SW93,W95}.
The presence of the above $f-u$ term leads to certain imaginary terms
in the 2PN accurate Kepler Equation
while adapting the usual argument of analytic continuation,
namely $ u \rightarrow \imath v$, to obtain its hyperbolic version.
An interesting extension will be to incorporate the effects of dominant order spin-spin interactions.
Another challenging direction will be to adapt Refs.~\cite{BD12,D14}
to describe GW burst signals while employing the framework of effective-one-body formalism.
It will be also desirable to pursue possible data analysis implications of these templates.
A possible direction may involve probing 
the ability of GW search algorithms like in Ref.~\cite{TC14}, constructed to capture   
unmodeled gravitational-wave bursts, to detect and distinguish our accurately modeled GW bursts.

\section*{Acknowledgments}
We would like to thank Simone Balmelli and C\'edric Huwyler for useful discussions, and
the SNF for hosting AG at the Physik-Institut in Z\"urich during the initial
stages of this work. We thank C. Berry for his detailed review of this article
and the referee for many helpful suggestions. This is a LIGO document, LIGO-P1400065.

\newpage

\onecolumngrid
\appendix

\section{1PN accurate polarization states expressions for spinning binaries}

\noindent We list below the 1PN accurate expressions for $h_+$ and $h_{\times}$ for spinning binaries on general orbits in a compact
form that incorporate 1PN accurate non-spinning and 1PN order spin-orbit contributions.
\begin{subequations}
\label{eq:hxp1PN}
\begin{align}
h_{+} &=  2\, \frac {G\, \mu  } {c^4\, R } \,  \bigg\{  \bigg[ \left( {(\bm{q} \cdot \bm{n} )}^{2}-{(\bm{p} \cdot \bm{n} )}^{2} \right) \,z +
	{(\bm{p} \cdot \textbf{v} )}^{2}- {(\bm{q} \cdot \textbf{v} )}^{2}  \bigg] -
	\frac{X_1 - X_2}{2\,c }\,  \bigg[  (  (\bm{N} \cdot \bm{n} )\, {\dot r} -(\bm{N} \cdot \textbf{v} )\,  )\, z\, {(\bm{p} \cdot \bm{n} )}^{2} \nonumber \\
&-6\,z\,(\bm{N} \cdot \bm{n} )\, (\bm{p} \cdot \bm{n} )\,(\bm{p} \cdot \textbf{v} )+ \left( -3\,(\bm{N} \cdot \bm{n} )\, {\dot r} +(\bm{N} \cdot \textbf{v} )\, \right) \,z\, {(\bm{q} \cdot \bm{n} )}^{ 2}
	+6\,z\,(\bm{N} \cdot \bm{n} )\, (\bm{q} \cdot \bm{n} )\,(\bm{q} \cdot \textbf{v} ) \nonumber \\
&+ 2 \left( {(\bm{p} \cdot \textbf{v} )}^{ 2}-\,{(\bm{q} \cdot \textbf{v} )}^{2} \right) (\bm{N} \cdot \textbf{v} )  \bigg]
	+ \frac{1}{6\,c^2}\,  \bigg[ 6\,{(\bm{N} \cdot \textbf{v} )}^{2}\,  (  (\bm{p} \cdot \textbf{v} )^2 - (\bm{q} \cdot \textbf{v} )^2   )
	\left( 1-3\,\eta \right) +  (  \left[ 6\,\eta-2 \right] {(\bm{N} \cdot \textbf{v} )}^{2}{(\bm{p} \cdot \bm{n} )}^{2} \nonumber \\
&+ \left( 96\,\eta-32 \right) \, (\bm{N} \cdot \textbf{v} ) \,(\bm{N} \cdot \bm{n} )\,(\bm{p} \cdot \textbf{v} )\,(\bm{p} \cdot \bm{n} )+ \left( -6\,\eta+2 \right) 
	\, (\bm{N} \cdot \textbf{v} )^2\, {(\bm{q} \cdot \bm{n} )}^{2} + \left( -96\,\eta+32 \right) (\bm{N} \cdot \textbf{v} )\, (\bm{N} \cdot \bm{n} ) \nonumber \\
&\times (\bm{q} \cdot \textbf{v} )\,(\bm{q} \cdot \bm{n} ) +  [   \left( -14+42\,\eta \right) (\bm{N} \cdot \bm{n} )^2 - 4+6\,\eta  ]  {(\bm{p} \cdot \textbf{v} )}^{2}+
	 [   \left( -42\, \eta+14 \right) {(\bm{N} \cdot \bm{n} )}^{2}+4-6\,\eta  ]  {(\bm{q} \cdot \textbf{v} )}^{2}  ) \, z \nonumber \\
&+  \left(  \left( -9\,\eta+3 \right) {(\bm{p} \cdot \textbf{v} )}^{2} + \left( - 3+9\,\eta \right) {(\bm{q} \cdot \textbf{v} )}^{2}  \right) 
	\,v^2 +  (   [  29+ \left( 7-21\,\eta \right) {(\bm{N} \cdot \bm{n} )}^{2}  ]  {(\bm{p} \cdot \bm{n} )}^{2} +  [ -29+ \left( 21\,\eta-7 \right) \nonumber \\
&\times {(\bm{N} \cdot \bm{n} )}^{2}  ] \, {(\bm{q} \cdot \bm{n} )}^{2}  ) {z}^{2} +  (  (  \left( -9\,\eta+3 \right) {(\bm{N} \cdot \bm{n} )}^{2}-10-3\,\eta )
	{(\bm{p} \cdot \bm{n} )}^{2}+ \left(  \left( -3+9\,\eta \right) {(\bm{N} \cdot \bm{n} )}^{2}+10+3\,\eta \right) \nonumber \\
&\times {(\bm{q} \cdot \bm{n} )}^{2} )\, z\, v^2 +  (  \left( -36\,\eta+12 \right) (\bm{N} \cdot \textbf{v} )\,(\bm{N} \cdot \bm{n} )\,{(\bm{p} \cdot \bm{n} )}^{2}+
	\left(  \left( -90\,\eta+30 \right) {(\bm{N} \cdot \bm{n} )}^{2}+20+12\,\eta \right) (\bm{p} \cdot \textbf{v} )\,(\bm{p} \cdot \bm{n} ) \nonumber \\
&+ \left( -12+36\,\eta \right) (\bm{N} \cdot \textbf{v} )\, (\bm{N} \cdot \bm{n} ) \,{(\bm{q} \cdot \bm{n} )}^{2}+ \left(  \left( 90\,\eta-30 \right) {(\bm{N} \cdot \bm{n} )}^{2}
	-12\,\eta-20 \right) (\bm{q} \cdot \textbf{v} )\,(\bm{q} \cdot \bm{n} )  ) \, z\, {\dot r} \nonumber \\
&+ \left(   [  \left( 45\,\eta-15 \right) {(\bm{N} \cdot \bm{n} )}^{2}-9\,\eta+3  ] \, {(\bm{p} \cdot \bm{n} )}^{2}+ \left(  \left( 15-45\,\eta \right) {(\bm{N} \cdot \bm{n} )}^{2}-3+9\,
	\eta \right) {(\bm{q} \cdot \bm{n} )}^{2}  \right)\,  \bigg] + \frac{ z^2  }{ c^2} \,  \bigg[ (\bm{p} \cdot \bm{n} )\nonumber \\
& \times  (X_2 \chi_2 \left [ \bm{p} \cdot \left ( {\bm{s}_2} \times  \bm{N} \right ) \right ] - X_1\chi_1\left [ 
	\bm{p} \cdot \left ( {\bm{s}_1} \times  \bm{N} \right ) \right ]  ) + (\bm{q} \cdot \bm{n} )  ( X_1 \chi_1 \left [
	\bm{q} \cdot \left ( {\bm{s}_1} \times  \bm{N} \right ) \right ] - X_2 \chi_2 \left [ \bm{q} \cdot \left ( {\bm{s}_2} \times  \bm{N} \right ) \right ]  )  \bigg] \bigg\}~, \\
h_{\times} &=4\, { \frac {G\, \mu  }{ {c^4\, R } } }\,  \bigg\{  \bigg[ -(\bm{p} \cdot \bm{n} )\,(\bm{q} \cdot \bm{n} )\,z +(\bm{p} \cdot \textbf{v} )\,(\bm{q} \cdot \textbf{v} )  \bigg] 
	- \frac{X_1 - X_2}{c} \, \bigg[  \bigg( \big\{  [  3\,(\bm{N} \cdot \bm{n} )\,{\dot r} -(\bm{N} \cdot \textbf{v}) ]  (\bm{q} \cdot \bm{n} )
	-3\,(\bm{N} \cdot \bm{n} ) \,(\bm{q} \cdot \textbf{v} ) \big\} \nonumber \\
&\times (\bm{p} \cdot \bm{n} ) -3\,(\bm{N} \cdot \bm{n} )\,(\bm{q} \cdot \bm{n} )\,(\bm{p} \cdot \textbf{v} ) \bigg)\, z +2\,(\bm{p} \cdot \textbf{v} )\,(\bm{q} \cdot \textbf{v} )\,(\bm{N} \cdot \textbf{v} )
	\bigg] + \frac{1}{6\, c^2}\,  \bigg[ 6\,  \left( 1-3\,\eta \right) {(\bm{N} \cdot \textbf{v} )}^{2}\, (\bm{p} \cdot \textbf{v})\,(\bm{q} \cdot \textbf{v} ) \nonumber \\
&+  (    [   \left( 6\,\eta-2 \right) (\bm{N} \cdot \textbf{v} )^{2}\, (\bm{q} \cdot \bm{n} )+ \left( 48\,\eta-16 \right)
	(\bm{N} \cdot \textbf{v} )\,(\bm{N} \cdot \bm{n} )\,(\bm{q} \cdot \textbf{v} )  ]  (\bm{p} \cdot \bm{n} ) + \left( 48\,\eta-16 \right) (\bm{N} \cdot \textbf{v} )\,(\bm{N} \cdot \bm{n} )\,
	(\bm{p} \cdot \textbf{v} )\,(\bm{q} \cdot \bm{n} ) \nonumber \\
&+ \left(  \left( -14+42\,\eta \right) {(\bm{N} \cdot \bm{n} )}^{2} -4+6\,\eta \right) (\bm{q} \cdot \textbf{v} )\,(\bm{p} \cdot \textbf{v} )  ) z
	+  ( -9\,\eta+3  ) (\bm{q} \cdot \textbf{v} )\,(\bm{p} \cdot \textbf{v} )\,v^2+  ( 29+ \left( 7-21\,\eta \right) {(\bm{N} \cdot \bm{n} )}^{2}  ) \nonumber \\
&\times (\bm{q} \cdot \bm{n} )\,(\bm{p} \cdot \bm{n} )\,{z}^{2}+  (  \left( -9\,\eta+3 \right) {(\bm{N} \cdot \bm{n} )}^{2}-10-3\,\eta  ) (\bm{q} \cdot \bm{n} )\,(\bm{p} \cdot \bm{n} )\,z\, v^2+
	 (  [   \left( -36\,\eta+12 \right) (\bm{N} \cdot \textbf{v} )\,(\bm{N} \cdot \bm{n} ) \,(\bm{q} \cdot \bm{n} ) \nonumber \\
&+ \left(  \left( 15-45\,\eta \right) {(\bm{N} \cdot \bm{n} )}^{2}+10+6\,\eta  \right) (\bm{q} \cdot \textbf{v} )  ] (\bm{p} \cdot \bm{n} ) +  [  \left( 15-45\,\eta
	\right) {(\bm{N} \cdot \bm{n} )}^{2}+10+6\,\eta  ] (\bm{p} \cdot \textbf{v} )\,(\bm{q} \cdot \bm{n} )  ) \dot r\,z +  (  \left( 45\,\eta-15 \right) \nonumber \\
&\times {(\bm{N} \cdot \bm{n} )}^{2}-9\,\eta+3  ) (\bm{q} \cdot \bm{n} )\,(\bm{p} \cdot \bm{n} )\,{\dot r}^{2}\,z  \bigg]  
	+ \frac{ z^2 }{ c^2} \, (\bm{q} \cdot \bm{n} )  \bigg[ X_2 \chi_2  ( \bm{p} \cdot \left ( {\bm{s}_2} \times  \bm{N} \right )  )
      - X_1 \chi_1 ( \bm{p} \cdot \left ( {\bm{s}_1} \times  \bm{N} \right )  )  \bigg]  \bigg\}~.
\end{align}
\end{subequations}
\noindent The dot products that appear in the above equations, as expected, are evaluated in the 
$(\bm{p},\bm{q},\bm{N})$ frame. These dot products can be 
written in terms of the Eulerian angles  $\Phi(t)$, $\alpha(t)$, $\iota(t)$, 
and the constant angle $\theta$ and are given by
\begin{align}
  \bm{p}\cdot\bm{n}&=-\cos\Phi\sin\alpha-\sin\Phi\cos\alpha\cos\iota~,\label{eq:pdotn}\\
  \bm{q}\cdot\bm{n}&=C_{\theta}(\cos\Phi\cos\alpha-\sin\Phi\sin\alpha\cos\iota)-S_{\theta}\sin\iota\sin\Phi~,\\
  \bm{p}\cdot\textbf{v}&=r\dot{\Phi}\,\left(\sin\Phi\sin\alpha-\cos\Phi\cos\alpha\cos\iota\right)-\dot{r}\,\left(\cos\Phi\sin\alpha+\sin\Phi\cos\alpha\cos\iota\right)~,\\
  \bm{q}\cdot\textbf{v}&=\dot{r}\,\left(C_{\theta}(\cos\Phi\cos\alpha-\sin\Phi\sin\alpha\cos\iota)-S_{\theta}\sin\Phi\sin\iota\right)\\
		       &-r\dot{\Phi}\,\left(C_{\theta}(\sin\Phi\cos\alpha+\cos\Phi\sin\alpha\cos\iota)+S_{\theta}\cos\Phi\sin\iota\right) ~,\nonumber\\
  \bm{N}\cdot\bm{n}&=C_{\theta}\sin\Phi\sin\iota+S_{\theta}(\cos\Phi\cos\alpha-\sin\Phi\sin\alpha\cos\iota)~,\\
  \bm{N}\cdot\textbf{v}&=\dot{r}\,\left(S_{\theta}(\cos\Phi\cos\alpha-\sin\Phi\sin\alpha\cos\iota)+C_{\theta}\sin\Phi\sin\iota  \right)\label{eq:Ndotv}\\
		       &-r\dot{\Phi}\,\left(S_{\theta}(\sin\Phi\cos\alpha+\cos\Phi\sin\alpha\cos\iota)-C_{\theta}\sin\Phi\sin\iota \right)~.\nonumber
\end{align}
The explicit evaluation of Eqs.~(\ref{eq:hxp1PN}) with the help of the above dot products leads to amplitude corrected GW polarization
states for spinning compact binaries on general orbits. We can numerically impose the evolution of various variables 
for the hyperbolic orbits 
to obtain the amplitude corrected polarization states for spinning binaries on hyperbolic orbits.

\twocolumngrid

\newpage

\bibliography{mybib}
\bibliographystyle{apsrev}

\end{document}